\documentclass[11pt,letterpaper]{article}
\usepackage[nolists, tablesfirst, nomarkers]{endfloat}
\usepackage{rotating}
\usepackage{natbib}
\usepackage{multirow}
\usepackage{amsthm,amsfonts,amsmath,scalerel }
\usepackage{graphicx}
\usepackage{enumitem}
\usepackage{subfigure}
\usepackage{capt-of}
\usepackage[font=scriptsize]{caption}
\usepackage{float}
\usepackage{tabularx}
\usepackage{tabulary}
\usepackage{soul}
\usepackage{color}
\RequirePackage[colorlinks,citecolor=blue,urlcolor=blue]{hyperref}
\usepackage[mathscr]{eucal}
\usepackage{url}

\def\UrlOrds{\do\*\do\-\do\~\do\'\do\"\do\-}%
\makeatletter
\g@addto@macro{\UrlBreaks}{\UrlOrds}
\makeatother

\newcolumntype{R}{>{\raggedleft\arraybackslash}X}
\newcolumntype{L}{>{\raggedright\arraybackslash}X}

\DeclareDelayedFloatFlavor{sidewaystable}{table} 
\DeclareDelayedFloatFlavor{sidewaysfigure}{figure}

\newcommand{\ci}{\citet}

\renewcommand{\labelenumi}{(\roman{enumi})}

\newcommand{\be}{\begin{equation}}
\newcommand{\ee}{\end{equation}}
\newcommand{\bew}{\begin{equation*}}
\newcommand{\eew}{\end{equation*}}

\definecolor{darkgreen}{RGB}{85, 166, 48}

\newfloat{fighere}{h}{lop}[section]

\allowdisplaybreaks

\begin{document}

\renewcommand{\labelenumi}{(\arabic{enumi})}
\renewcommand{\baselinestretch}{1.5} \small \normalsize

\title{\LARGE{Preventing COVID-19 Fatalities:  State versus Federal Policies}}

\author{J.-P. Renne, \; G. Roussellet,\; and \; G. Schwenkler\footnote{Renne is at the Department of Economics of the University of Lausanne. Roussellet is at the Desautels School of Management of McGill University. Schwenkler is at the Leavey School of Business of Santa Clara University. Schwenkler is the corresponding author. Address: Leavey School of Business, Santa Clara University, 500 El Camino Real, Santa Clara, CA 95053, United States. Email: gschwenkler@scu.edu.} \\  }

\date{First version: October 27, 2020 \\ Current version: \today\footnote{We are grateful to Patrick Augustin, Vadim Elenev, Christian Gouri\'eroux, Andrea Vedolin, Charles Wyplosz, and seminar participants at Santa Clara University and at HEC-Unisant\'e COVID Seminar for useful comments and suggestions. The latest version of the codes used in this paper can be found at \url{https://github.com/guillaume-roussellet/RRS_Covid_2020}.}}
\maketitle


\begin{abstract}
	Are COVID-19 fatalities large when a federal government does not enforce containment policies and instead allow states to implement their own policies?
We answer this question by developing a stochastic extension of a SIRD epidemiological model for a country composed of multiple states. Our model allows for interstate mobility. We consider three policies: mask mandates, stay-at-home orders, and interstate travel bans. We fit our model to daily U.S. state-level COVID-19 death counts and exploit our estimates to produce various policy counterfactuals. While the restrictions imposed by some states inhibited a significant number of virus deaths, we find that more than two-thirds of U.S. COVID-19 deaths could have been prevented by late November 2020 had the federal government enforced federal mandates as early as some of the earliest states did. 
Our results quantify the benefits of early actions by a federal government for the containment of a pandemic.

\vspace{1cm}

\end{abstract}

\thispagestyle{empty}
\newpage

\section{Introduction} 

COVID-19 is a rampant disease that has affected the world population to an unprecedented scale. This experience has pushed governments to implement drastic regulatory policies to contain the disease. In many countries, such as the United States, tensions have arisen between federal and state regulators because they can implement policies independently of each other. But are policies implemented in individual states effective to prevent COVID-19 fatalities in a country? And can a country's defense against COVID-19 benefit from unified containment policies at the federal level? The answer to these questions have important implications for policy making until the end of the pandemic.

We answer these questions by developing an extension of the standard SIRD model of \ci{kermack-et-al-1927} that allows for travel and commuting across states within a country.\footnote{We are inspired by mean-field models that are commonly used in the financial economics literature to model credit risk contagion across financial institutions; see \ci{cvitanic-et-al-2012}, \ci{LLN}, \ci{largesystems}, and others. Our model can be viewed as an extension of a global epidemic and mobility model \citep[GLEAM, see][]{balcan-et-al-2009} with stochastic transmission rates.}
In our model, the government of a state can enforce three types of regulations. It can enforce a mask mandate that shrinks the transmission rate in the state. It can enforce a stay-at-home order that shrinks the transmission rate in the state as well as the inflow of out-of-state commuters and travelers. 
Or it can also issue a travel ban that shrinks the inflow of out-of-state travelers. The federal government can force all states to implement the same policies or do nothing. We assume that infections take on average 14 days to resolve and that the fatality rate of the disease is 0.6\%.\footnote{These assumptions are consistent with \ci{Fernandez-Villaverde_Jones_2020}, \cite{PerezSaez_et_al_2020}, and \cite{Stringhini_et_al_2020}, and are benchmarked against alternatives in a sensitivity study.}
To incorporate uncertainty about the contagiousness of COVID-19, we assume that the transmission rates in individual states vary randomly over time and are not directly observable.
They fluctuate around a natural mean rate but can be substantially higher or lower at times.\footnote{Many different variations of the SIRD model have been used to study the COVID-19 pandemic; see \cite{LI2021130} for a recent overview.} 
We fit our model to data on state-level COVID-19 fatalities from the United States between February 12 and November 30, 2020. We then run counterfactual experiments using the estimated model under the assumption that states implemented policies different than the ones they adopted.
We measure the effectiveness of the different policies by looking at the difference between the observed and counterfactual numbers of virus deaths on November 30, 2020.\footnote{Our counterfactuals do not consider that changes in policies may have triggered different strategies in the states and different reactions by the U.S. population. A novel approach to account for this endogeneity problem has been proposed by \ci{chernozhukov-et-al-2020}.} 
Our results are replicable by using our codes and data that are available on \href{https://github.com/guillaume-roussellet/RRS_Covid_2020}{Github}.

We show that a lack of unified policies results in significantly elevated virus deaths nationally.
 We estimate that more than 180,000 deaths could have been prevented by November 30, 2020 -- over two-thirds of all death cases recorded in the U.S. by that date -- if the federal government had adopted policies that mirrored those of the earliest and strictest states.
Our results also show that containment policies implemented by individual states are effective. We find that the U.S. would have recorded more than 1 million additional virus deaths if states had not implemented any containment policies at all. 

We find that a large number of  COVID-19 deaths could have been prevented if the federal government had enacted policies that followed the leads taken by the different states.\footnote{In our model, a federal mandate is equivalent to all states voluntarily adopting that mandate. As a result, our paper equally implies that a significant number of deaths could have been prevented if the federal government had convinced all states to adopt policies that mirrored those of the earliest states.}
We estimate that close to 130,000 deaths could have been prevented if a federal stay-at-home order had been active between March 19 and June 9, 2020. This period corresponds to the time between the day when the first stay-at-home order went into effect in California and the day when Virginia began to reopen its economy as the last state in the country. More than 140,000 virus deaths could have been prevented with a federal mask mandate that became active on April 17 and remained active through the end of our sample, mirroring Connecticut's policy as the earliest state to adopt a mask mandate. Only around 8,000 U.S. virus death could have been prevented if all interstate travel had been banned beginning on March 17, reflecting the travel restrictions adopted by Hawaii. 


Our results suggest that a similar number of COVID-19 deaths could have been prevented in the United States if the federal government had followed the early states' leads on mask mandates or lockdowns. However, this conclusion ignores the distinct timing of both policies: state stay-at-home policies were enacted earlier and for a shorter periods of time than state mask mandates. We run an additional counterfactual to understand how an early federal mask mandate could have boosted the early stay-at-home orders adopted by the different states.

Our counterfactuals indicate that up to 235,000 virus deaths could have been prevented if the federal government had enacted a federal mask mandate by March 19, on top of the stay-at-home and travel ban policies adopted by the different states. This finding suggests that over 90\% of all COVID-19 fatalities in the U.S. could have been prevented if the federal government had enforced an early federal mask mandate and allowed states to implement lockdowns and travel restrictions as they chose to. 
This is a surprising result that holds because an early federal mask mandate would have contributed to slowing down transmission of the virus early in the Spring of 2020, when we estimate the virus reproduction numbers in the different states to have been the highest.\footnote{By matching death counts only, we estimate that the effective reproduction numbers of the virus in the different states during the Spring of 2020 must have been up to two-times higher than prevailing estimates that are based on infection cases, which are likely under-measured due to the large number of undetected infections and asymptomatic individuals.}
These results provide an important policy lesson for federal regulators: a federal mask mandate that complements state-level lockdowns and travel restriction during periods of accelerated virus transmission can prevent a significant number of COVID-19 deaths.\footnote{Our results highlight one effective policy strategy for federal regulators but they do not say anything about the optimality of such a strategy. Furthermore, we do not consider the legality of any federal mandate.}

We find that interstate travel bans do not accomplish significant virus death prevention, mostly for two reasons. First, travel bans are often imposed once a regulator becomes aware of the virus, at which point the virus has  already spread in the states. We find that an additional 1,300 deaths could have been prevented if all interstate travel had been banned by February 12, the first day of our sample, rather than by March 17, the day in which Hawaii adopted the first interstate travel restrictions in the United States.

Second, there exists a mechanism that can reduce the effectiveness of travel restrictions in some states. Our model highlights that restrictions on interstate mobility can increase the concentration of infected individuals in states that are normally net exporters of individuals. This can lead to an accelerated virus transmission and elevated virus fatalities in those states. Our counterfactuals suggest that such contrary effects are only minor, nonetheless. We find that only around 1,700 of the close to 130,000 virus deaths that could have been prevented in the U.S. with a federal stay-at-home order by March 19 are explained by the fact that the order discourages individuals from commuting and traveling across states.
Our results suggest that policies that restrict cross-border mobility accomplish little in preventing COVID-19 deaths unless they are adopted so early and for so long that they prevent the virus from initially taking hold in a population.\footnote{\ci{kortessis-et-al-2020} also document the advantages of early travel bans. For the U.S. in specific, an early interstate travel ban could have prevented the urban flight phenomenon highlighted by \ci{coven-et-al-2020}.}

Finally, we focus on individual states and find that the states that adopted some of the strictest containment policies were able to reduce the spread of the disease. Our counterfactuals suggest that New York would have only recorded 20\% fewer death cases if all states had adopted strict containment policies, while California would have recorded close to 145,000 additional deaths if the state government had not enforced any containment policies at all.
These results suggest that strict-policy states, such as New York and California, were protected by their policies even when other states implemented weaker policies.
On other end of the spectrum, we find that states that adopted weak or no containment policies could have prevented a significant number of COVID-19 deaths by adopting stricter policies. Our results indicate that over 38,000 deaths could have been prevented if Florida and Texas had adopted strict containment policies. We also find that the twelve states that never adopted a mask mandate over our sample -- Alaska, Arizona, Florida, Georgia, Idaho, Missouri, Nebraska, Oklahoma, South Carolina, South Dakota, Tennessee, and Wyoming -- could have prevented more than 76\% of their death cases if they had followed Connecticut's lead and adopted a mask mandate by April 17, 2020. These results highlight an important takeaway for state regulators: state mask mandates are highly effective even in the absence of federal policies.

Our results hinge on modeling choices and assumed parameter values. To understand how sensitive our findings are to our choices, we carry out several sensitivity analyses that assume alternative parametric values and provide a reasonable set of bounds for our results. We find, for example, that the number of preventable deaths is higher if we assume that it takes less times for a virus infection to resolve; i.e., if the virus is less severe. This is because, if a virus infection is less severe, there must have been many more infections early on in the sample to match the observed death data. Therefore, early action by the federal regulator would have been even more impactful. 
The sensitivity analyses highlight a key benefit of our approach. By relying on death counts only, we allow our methodology to infer how many infections there must have been to justify the death data. This enables us to make data-driven inferences on how many infections and ultimate deaths could have been prevented through regulatory actions.

Our contributions are methodological and normative. On the methodological front, we develop a novel model for infectious disease transmission within a country composed of multiple states and propose an associated estimation methodology. Our model incorporates the effects of interstate traveling and commuting, as well as uncertain transmission rates.\footnote{Stochastic SIRD models of COVID-19 have recently been used by \ci{Fernandez-Villaverde_Jones_2020} and \ci{hong-et-al-2020}. Interstate SIRD models without stochastic transmission rates have been proposed by \ci{read-keeling-2003}, \ci{balcan-et-al-2009}, \ci{balcan-et-al-2010}, and others.} 
It also allows for independent regulatory policies across states.\footnote{\ci{brady-et-al-2020} study an SIRD model with spatial interactions and independent stay-at-home policies across states, without considering mask mandates and travel restrictions.}
Our methodology delivers daily estimates of the transmission rates and effective reproduction numbers; these objects are jointly estimated for 51 U.S. states based solely on death cases.\footnote{For single-state models, a related methodology is discussed in \cite{gourieroux-jasiak-2020} and implemented in \cite{ArroyoMarioli_et_al_2020}, \cite{Hasan_et_al_2020}, and \cite{Hasan_Nasution_2020} using infections data. \cite{Fernandez-Villaverde_Jones_2020} develop an alternative filtering-based methodology to obtain time-varying estimates of the effective reproduction number $\mathcal{R}_t$ using death counts only for a model without traveling or commuting.}
On the normative side, we show that a federal mask mandate that complements state-level lockdowns and travel restrictions can prevent a significant number of COVID-19 fatalities in a country.
Our results hint that close to 180,000 coronavirus deaths could have been prevented if the federal government in the United States had followed the lead of some of the states that took early actions to contain the virus.\footnote{\ci{redlener-et-al-2020} establish a similar number of preventable deaths through a comparison of international policy responses. As \ci{shefrin-2020} argues, however, cultural and ideological differences may have prevented the U.S. government from adopting international policies. We show that a large number of COVID-19 deaths could have been prevented by following policies that were already implemented domestically in the states. }
We also show that state mask mandates are highly effective, while travel restrictions have small effects at the federal level.\footnote{Our results are consistent with the findings of \ci{chernozhukov-et-al-2020}, \ci{karaivanov-et-al-2020}, \ci{ngonghala-et-al-2020}, and \ci{stutt-et-al-2020}, who highlight the benefits of a federal mask mandate in single-state models. 
Other studies that consider the impact of different federal containment policies include \ci{alfaro-et-al-2020}, \ci{alvarez-et-al-2020}, \ci{ferguson-et-al-2020}, \ci{Flaxman_et_al_2020}, and \ci{fowler-et-al-2020}, among others.} 
Our results highlight the benefits of early actions by federal and state regulators for containing pandemics.


\section{Model description}

We assume that a country is composed of several states. We model the number of people in each state that are \emph{(i)} susceptible to the virus (i.e., have never been infected), \emph{(ii)} infected by the virus, \emph{(iii)} recovered from the virus (and immune to subsequent infections), and \emph{(iv)} deceased due to the virus.\footnote{Our assumption on immunity post-infection is merely a simplification given that current evidence on re-infections from COVID-19 is thin and mixed \citep[see, e.g.][]{Iwasaki-Lancet-2020}.} 
Because of this, our model can be viewed as a multi-region SIRD model.
Extending the standard SIRD formulation of \ci{kermack-et-al-1927}, we allow for stochastic and correlated transmission rates in each state. We also explicitly take into account the impact of different containment policies on the transmission of the disease. We consider three different types of containment policies: mask mandates, travel bans, and stay-at-home orders. Our model also accounts for the effects of traveling and commuting across state lines.
We give a brief description of our model below and provide details in Appendix \ref{app:Model_Formulation}.





Each state is endowed with a certain number of inhabitants. On any given day in a given state, the number of people that are contaminated by an infected person is assumed to be a random draw from a Poisson distribution whose rate is the product of the policy-adjusted stochastic transmission rate in the state and the fraction of the state’s population that is susceptible to the virus.\footnote{It is our assumption that there are two sources of randomness driving infections in a state. First, the transmission rate is stochastic. Second, the number of infections given the level of the transmission rate is also stochastic and governed by a Poisson distribution. Such doubly-stochasticity improves the capacity of our model to capture the early period of the COVID-19 pandemic during which the transmission rate of the virus may have been high but the number of infections and fatalities in the data were low. A similar point for the avian influenza pandemic was made by \cite{Germann_et_al_2006}.}
Recovering and dying from a virus infection are modeled as complementary random events for each infected individual.
We assume that it takes on average 14 days for an infection to resolve and that the fatality rate by the end of this time period is 0.6\%.
On aggregate at the state level, it follows that the net number of new infections in a day is Poisson distributed with rate equal to the product of the policy-adjusted transmission rate, the ratio of susceptible population, and the number of infected inhabitants of the state, after subtracting the number of infected individuals that recover or die from the disease.

The transmission rates in the different states are unobservable and evolve stochastically from day to day. They are positively correlated across states and time. 



We account for interstate commuting and traveling as follows. Each day, an inhabitant of a state can commute to another state, travel to another state, or remain in the home state with given probabilities. 
Inhabitants that stay in their home states contribute to the disease transmission in their home states.
Commuters travel during parts of the day and return to their home states at the end of a day. They contribute to the transmission of the disease in their home states and the visited states. 
Travelers spend several days in the visited state. They only contribute to the 
transmission of the disease in the state they visit.

The transmission rates in each state, as well as the travel and commuting probabilities across states, are adjusted to reflect the containment policies adopted by the different states. 
The effects of the different policies are modeled by shrinking the transmission rates or the inflows of travelers and commuters when the policies are active. We assume that a mask mandate in a given state shrinks the transmission rate in that state by 42\% but does not affect the inflow or outflow of travelers and commuters.\footnote{We justify this value as follows. Based on the estimates of \ci{fischer-et-al-2020}, we assume that a typical mask worn in the U.S. filters out particles by a median value of 85\%. We assume that only half of the population wears masks as suggested by data from the Institute of Health Metrics and Evaluation, justifying a reduction factor of approximately 42\%.} 
A travel ban in a given state shrinks the inflow of travelers by 90\%, but does not affect the transmission rate or the inflow of commuters in that state.\footnote{We estimate the 90\% impact of travel bans as the reduction factor in the average number of daily trips of more than 100 miles that were taken in the U.S. between March 15 and April 30, 2020, compared to 2019. We assume that travel bans do not reduce the inflow of commuters into a state because travel bans were often accompanied by exemptions for commuters.} 
Finally, a stay-at-home order shrinks both the transmission rate and the inflow of commuters and travelers in that state by 36\%.\footnote{We estimate the impact of stay-at-home orders on transmission rates as the reduction factor in the average number of daily trips of less than 10 miles taken in 2019 versus those taken between March 15 and April 30, 2020, which is the period in which most stay-at-home orders were active in the United States. Because stay-at-home orders were often accompanied by lockdowns which closed out office buildings, we assume that stay-at-home orders also reduce the inflow of commuters into a state. Finally, because stay-at-home orders were also associated with closed out tourist attractions, we also assume that stay-at-home orders reduce the inflow of travelers into a state.} Appendix \ref{app:sens} considers variations of these parameters and measures the sensitivity of our results to our assumptions.



We use mobility, travel, and state-level policy data from the United States to calibrate most of the parameters of our model. We develop a filtering-based maximum likelihood methodology to estimate some of the parameters governing the transmission rates of the different states.
We rely on daily state-level death counts from The COVID Tracking Project since they are the least likely to be contaminated with substantial measurement issues (compared to the number of infections, for instance).\footnote{The fact that our estimation approach relies on death counts data only, as in \cite{Flaxman_et_al_2020}, is an advantage over standard approaches to estimate time-varying reproduction numbers (e.g. \cite{Bettencourt_Ribeiro_2008}, \cite{Cori_et_al_2013}, \cite{Thompson_2019}), as these approaches require data on infection cases. We also escape the identification problem noted by \cite{Korolev_2020} since we calibrate the parameters associated with the standard SIR model (recovery and death probability).}
The sample period is February 12 through November 30, 2020. 
Appendix \ref{sec:data} provides details of the approach we take to fit our model to the data. Figure \ref{fig:death} in Appendix \ref{sec:data} shows the death counts in the data as well as the posterior mean of the death counts in our model. The figure shows that our model performs well at matching state-level and aggregate death counts.

Figure \ref{fig:inf_r0} shows that, in order to match the levels of fatalities we observed over the sample period in U.S. states, our model estimates that the effective reproduction numbers of the disease in the U.S. must have been substantially higher than one through May 2020 (we break down the effective reproduction rate per state in Appendix \ref{sec:data}). 
The figure also shows that the effective reproduction rate grew larger than one several times over the summer and has remained consistently above $1$ since October, 2020. Our estimates suggest that the number of infections required to match the history of COVID-19 fatalities in U.S. states must have been substantially higher than recorded in the data.
We estimate that there must have been close to 48 million infections in the U.S. by November 30, almost four-times as many infections as in our data sources. Our estimates imply that many infected people remained undiagnosed in the U.S. and contributed to the spread of the disease, especially early during our sample (see \ci{bajema-et-al-2020}, \ci{Hortacsu_Liu_Schwieg_2020}, and others).

\section{Results}

We run counterfactual experiments in our model to study the effectiveness of the different containment policies. The underlying assumption of our counterfactuals is that states deviate from the policies they enacted by imposing stricter or looser restrictions in a hypothetical world. We ask: what would have happened to the death count of a state if it had enacted a stricter containment policy than it actually did? What would have happened if it had adopted a policy that it did not enact? And how would these deviations have impacted the aggregate death count in the U.S. by the end of our sample?
We answer these questions for mask mandates, stay-at-home orders, and travel ban policies. 

We assume that a state adopts a \emph{loose} policy if it does not impose that policy at all over our sample period. In contrast, we assume that a state adopts a \emph{strict} policy if it adopts that policy as early as the earliest state adopted the policy in the data, and keeps the policy active until the last state in our sample shuts off that policy. Our strict policy scenario therefore avoids forward-looking biases. Any regulator could have adopted a strict policy scenario in real life by moving along with the first state that adopted a policy and ending the policy as soon as no other state had the policy in place.\footnote{The legality of such federal mandates are  outside of the scope of this paper.} 
The strict scenarios for the different policies are:
	\begin{itemize}
		\item \textbf{Stay-at-Home}: Start on March 19 (the day the first stay-at-home order in a state went into effect in California) and end on June 9, 2020 (the day the last stay-at-home order in a state ended in Virginia).
		\item \textbf{Travel ban}: Start on March 17, 2020 (the day the first interstate travel restrictions in the U.S. went into effect in Hawaii) and keep it active through November 30, 2020 (when several states had travel restrictions in place).
		\item \textbf{Mask mandate}: Start on April 17, 2020 (the day the first mask mandate in the U.S. went into effect in Connecticut) and keep it active through November 30, 2020 (when several states had mask mandates in place).
	\end{itemize}	 

We consider two variations of our counterfactual experiments. One in which only one state at a time deviates from its actual policies at a time, and another one in which all states take on the same policy jointly at the same time. With the first set of experiments we seek to answer how impactful deviations at the state level would have been. The second set of counterfactuals studies how impactful federally mandated polices would have been. To evaluate the effectiveness of the different policies, we compare the number of deaths at the state and federal levels in the different counterfactual experiments to the baseline levels in the data. Appendix \ref{app:cf} provides details of our approach to computing counterfactual results.

Table \ref{tbl:fed} shows the results of the counterfactual experiments in which we assume that all states deviate jointly. Figure \ref{fig:federal} breaks down the number of deaths that could have been prevented through federal mandates by states and displays these in proportion to the number of death in the data of the individual states. Figure \ref{fig:state} shows the number of deaths that could have been prevented in the different states if individual states had deviated from their implemented policies while all other states kept their policies untouched. 

We carry out several robustness checks that re-evaluate the number of preventable deaths in the different scenarios under alternative assumptions of the impact of the different policies on transmission rates and traveler and commuter flows, as well as the fatality rate of the disease and the number of days it takes for an infection to resolve. The results of these sensitivity analyses are summarized in Table \ref{tbl:fed} and further elaborated in Appendix \ref{app:sens}. The additional sensitivity analyses are consistent with our main conclusions and provide validity for our results. 

\subsection{Federal mandates} \label{sec:fed}

Table \ref{tbl:fed} suggests that 180,595 virus deaths could have been prevented if the U.S. government had mandated federal containment policies as early as the earliest state did and maintained the policies for as long as the strictest states did. Our results indicate that 129,202 deaths  could have been prevented had the federal government moved along with California and mandated a federal stay-at-home order on March 19 that ended on June 9 when the last state, Virginia, reopened its economy. The results also indicate that 141,048 deaths could have been prevented if the federal government had issued a federal mask mandate as early as Connecticut did on April 17, 2020, and kept the mandate active through November 30. In contrast, we find that only 8,149 deaths could have been prevented had the federal government banned all interstate travel by March 17, 2020 -- the day in which Hawaii banned inbound interstate travel. 

\begin{sidewaysfigure}[t]
\begin{center}
\subfigure[All states impose strict or loose policies.]{ 
	\includegraphics[width=0.45\textwidth]{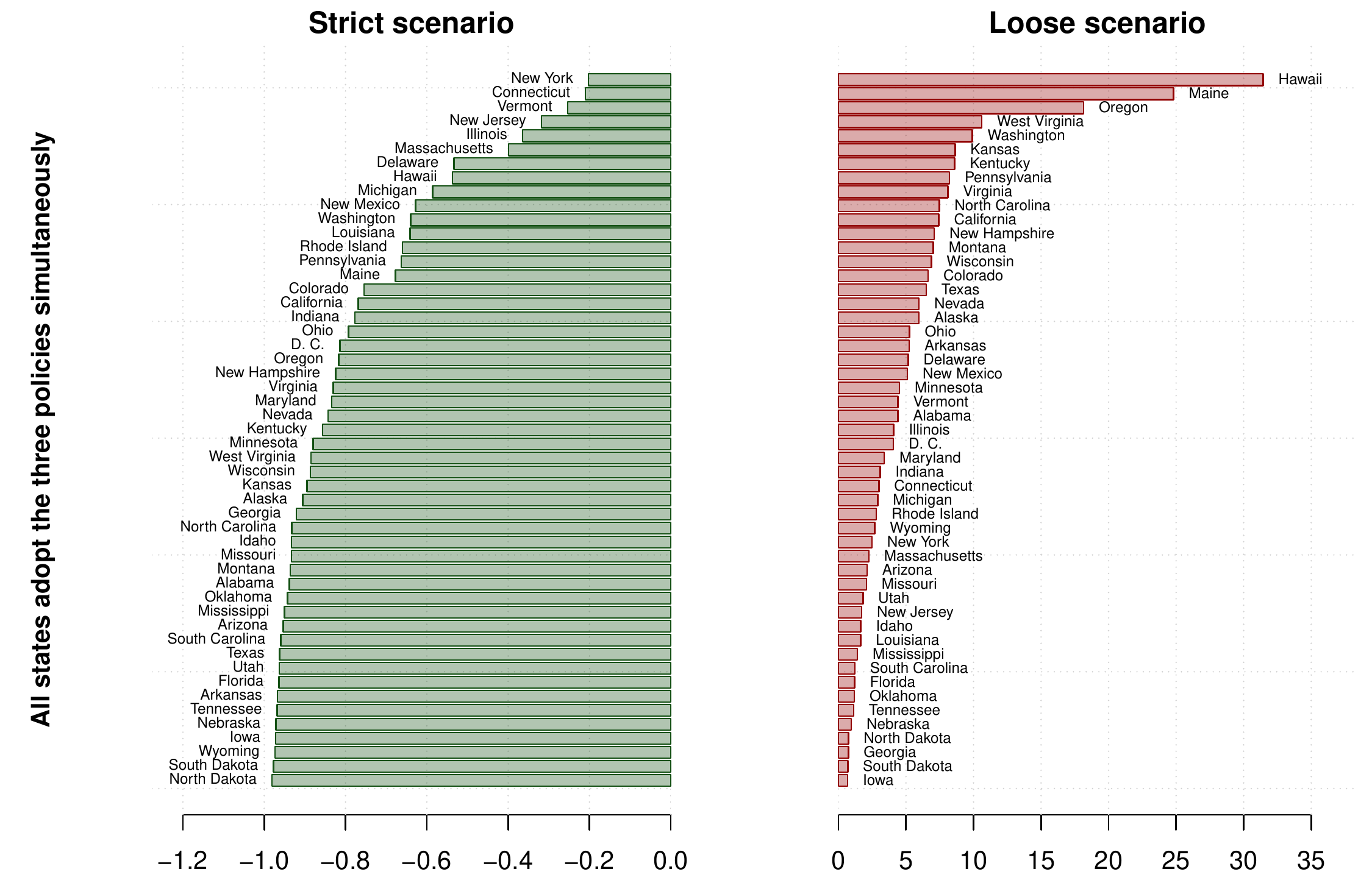}
} \qquad
\subfigure[All states impose strict or loose stay-at-home orders.]{ 
	\includegraphics[width=0.45\textwidth]{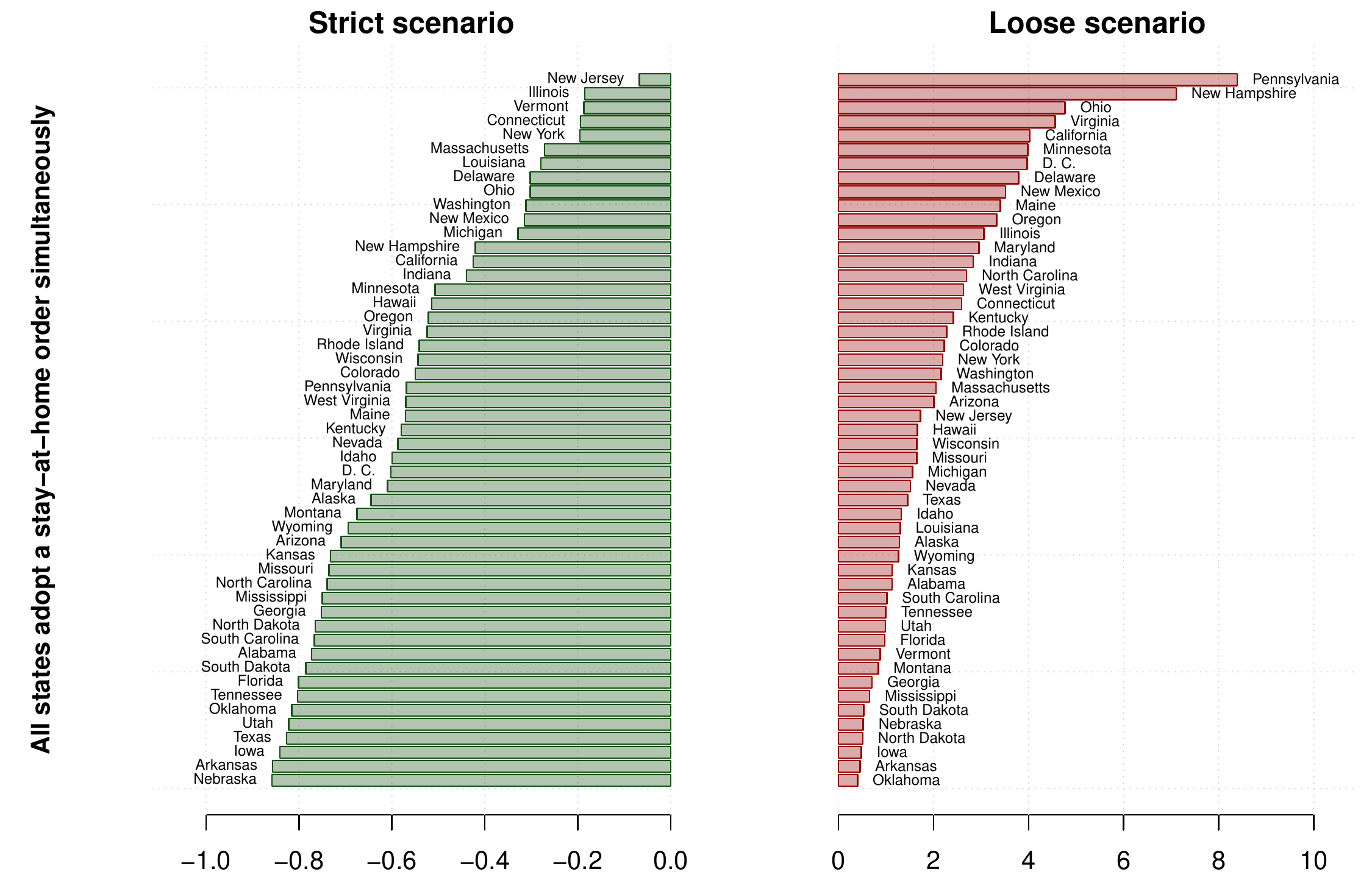}
} \qquad
\subfigure[All states impose strict or loose mask mandates.]{ 
	\includegraphics[width=0.45\textwidth]{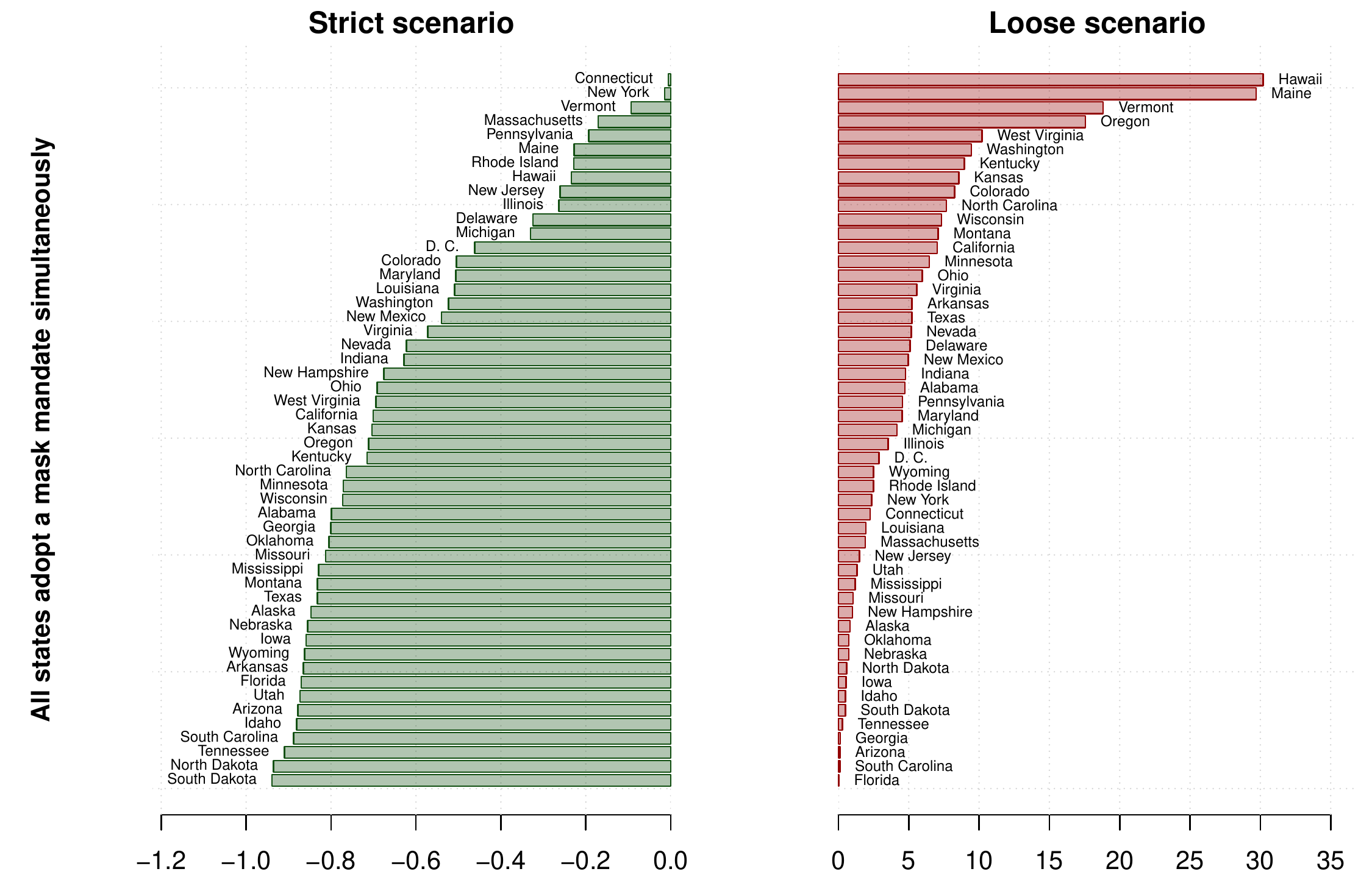}
} \qquad
\subfigure[All states impose strict or loose interstate travel bans.]{
	\includegraphics[width=0.45\textwidth]{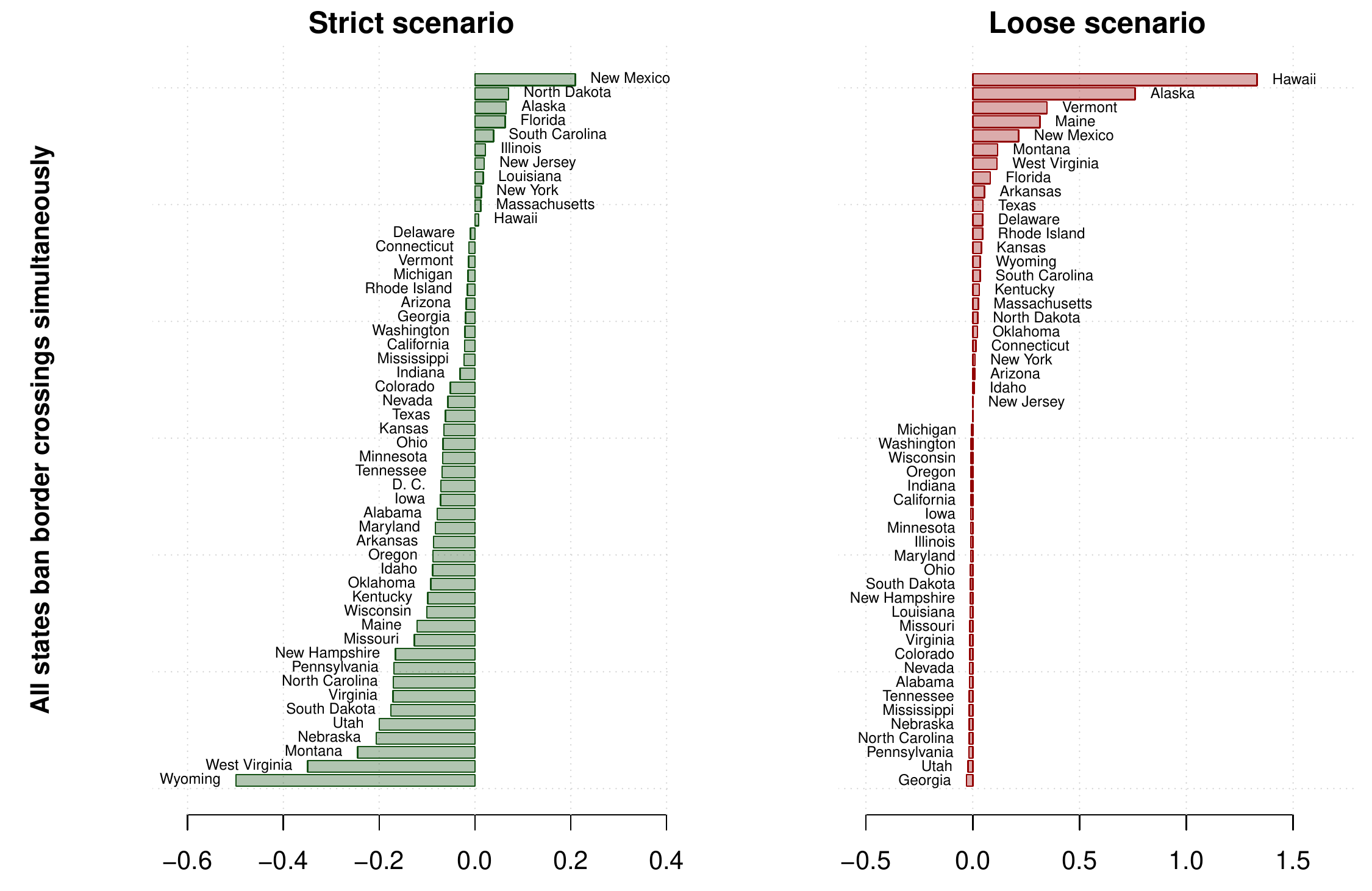}
}
\end{center}
\caption{State-by-state breakdown of excess deaths relative to the baseline in the different counterfactual scenarios in which all states jointly deviate from their implemented policies. The excess death values are measured as proportions of the death cases in the data for each state. Policies are divided into stay-at-home order (b), mask mandates (c), travel bans (d), and all three policies together (a). Counterfactual policies are divided into a \textit{strict} scenario in which all states implement a policy as long as at least one state decides to do so, and a \textit{loose} scenario in which no state implements a particular policy. Counterfactual death counts are computed through the methodology detailed in Appendix \ref{app:cf}.} 
\label{fig:federal}
\end{sidewaysfigure}

Figure \ref{fig:federal} hints that strict federal mandates could have been highly effective at preventing virus deaths in states that adopted weak containment policies. Over 95\% of all death cases in Arkansa, Arizona, Florida, Iowa, Nebraska, North Dakota, South Carolina, South Dakota, Tennessee, Texas, Utah, and Wyoming -- corresponding to close to 63,000 fatalities -- could have been prevented if the federal government had adopted strict stay-at-home orders, mask mandates, and interstate travel bans. These states adopted some of the weakest containment policies during our sample period.

\begin{figure}[tp]
\centering
\includegraphics[width = 0.9\textwidth]{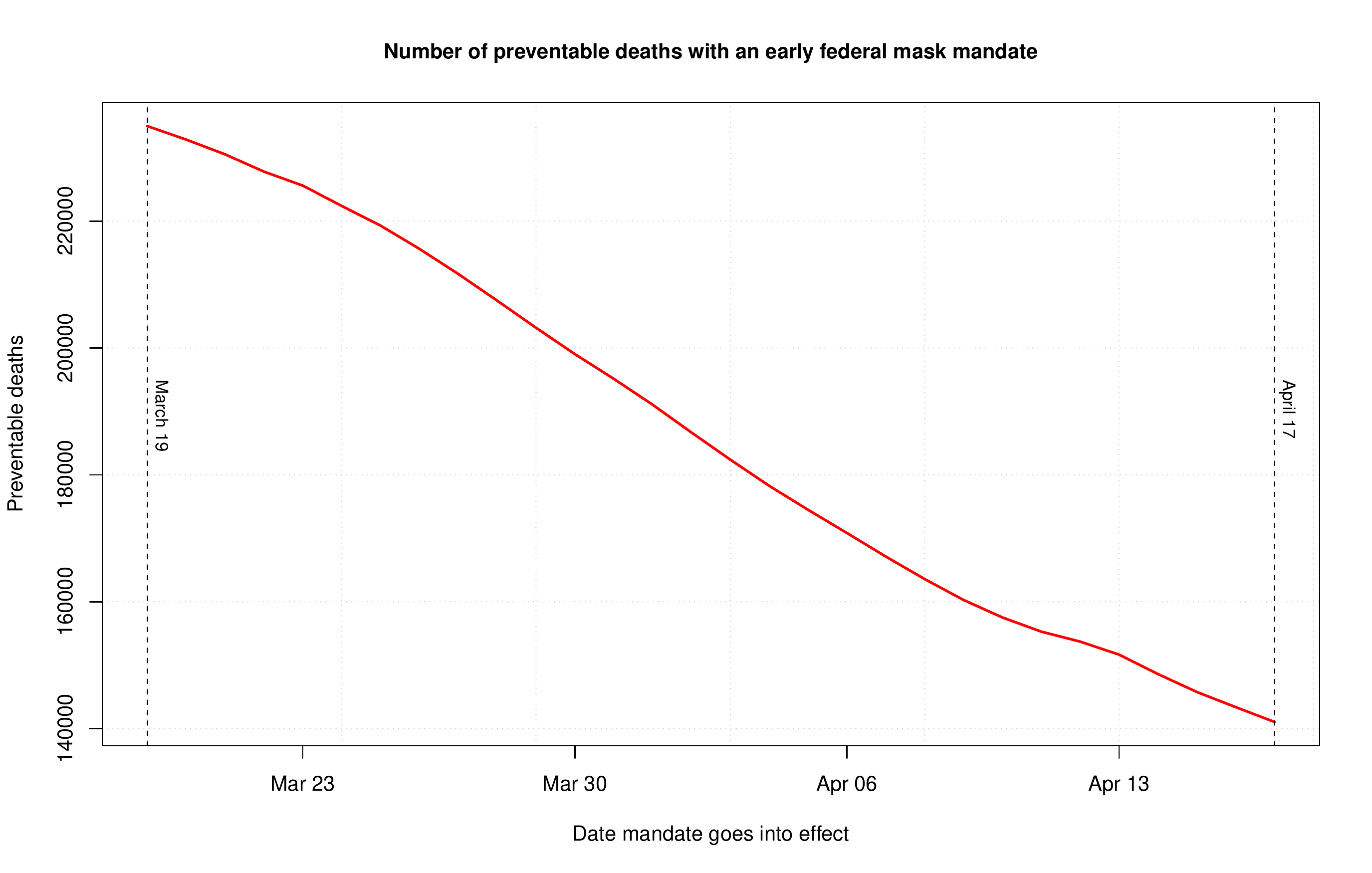} 
\caption{Number of deaths that could have been prevented through an early federal mask mandate. The $x$-axis indicates the date in which we assume a federal mask mandate had gone into effect, while the $y$-axis gives the number of deaths that could have been prevented had a federal mask mandate gone into effect on that date. We assume that state-level stay-at-home and travel ban policies remain as in the data, and that a federal mask mandate supersedes the state-level mask policies. }
\label{fig:masks}
\end{figure}

Table \ref{tbl:fed} suggests that a similar number of virus deaths could have been prevented with a federal mask mandate as with a federal lockdown. Such a conclusion, however, ignores the fact that we assume that a federal lockdown would become active earlier than a federal mask mandate. It also ignores the large economic costs that are associated with shutting down the national economy through a federal lockdown.\footnote{\ci{coibon-et-al-2020} measure the economic costs associated with lockdown policies.}
Imposing an early federal mask mandate could have been a cost-effective alternative. We ask: how many deaths could have been prevented if the federal government had enacted a federal mask mandate sometime between March 19 and April 17 and allowed states to adopt localized stay-at-home orders and travel restrictions as they chose to do in reality? We run an additional counterfactual experiment to answer this question; Figure \ref{fig:masks} summarizes our findings.

\begin{figure}[tp]
\centering
\includegraphics[width = \textwidth]{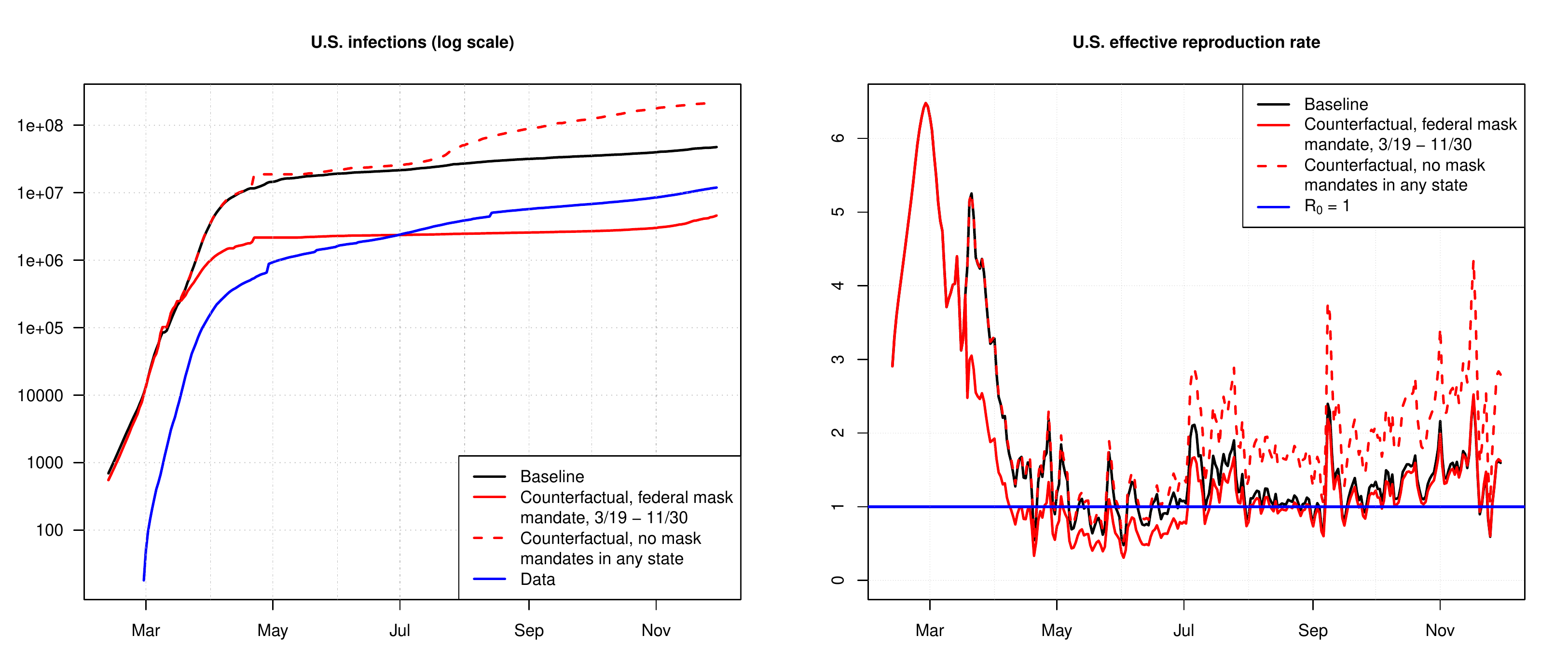} 
\caption{Cumulated infections implied by the model and the data, together with the estimated effective reproduction rate for the United States. The effective reproduction rate measure the number of susceptible individuals that an infected individual infects on average (see Appendix\;\ref{App:EffR0}). The rate varies over time in our model. The red lines correspond to the results a counterfactual analysis in which we assume that a federal mask mandate went into effect on March 19 and remained active through November 30, 2020; see Section \ref{sec:fed}.}
\label{fig:inf_r0}
\end{figure}

Our results indicate that between 141,048 and 236,551 deaths could have been prevented if a federal mask mandate that complemented the state-level stay-at-home and travel ban policies went into effect sometime between mid March and mid April.
What drives our findings is that an early federal mask mandate could have provided a significant boost in reducing the potential for infections across U.S. states. As Figure \ref{fig:inf_r0} shows, the effective reproduction number of the virus was particularly high -- much higher than one -- in the U.S. through May 2020 even while stay-at-home orders were in place in the different states. By imposing an early mask mandate, the federal regulator could have contributed to drastically reducing the effective reproduction rate of the virus early in the sample; see the solid red line on the right-hand plot of Figure \ref{fig:inf_r0}. This would have contributed to slowing down new infections over time even while some state-level policies were relaxed; see the solid red line in the left-hand plot of Figure \ref{fig:inf_r0}. 
Our results suggest that an effective tool for federal regulators is to issue an early federal mask mandate that complements, but does not replace, state-level stay-at-home and travel ban policies.

The number of deaths that could have been prevented with early federal policies depends on our assumptions on how fatal COVID-19 is and how long it takes for an infection to resolve. We run sensitivity analyses in Appendix \ref{app:sens} to benchmark our results against different calibrations. 
The results of Appendix \ref{app:sens} are generally consistent with Table \ref{tbl:fed} and Figures \ref{fig:federal} through \ref{fig:masks}. They also highlight an important benefit of our methodology. Appendix \ref{app:sens} establishes that the number of preventable deaths is higher if we assume that it takes only 10 days for an infection to resolve.\footnote{We find that the fatality rate has little impact on the number of preventable deaths, and instead mostly affects the number of deaths that would have been observed if no state adopted any policy.}
We find that this is the case because, in order to match the number of deaths observed in the data if the virus were less severe, the methodology infers that there must have been many more infected individuals early in the sample so that early federal action would have been even more impactful. We obtain these results because we rely only on death counts to make our inferences and allow the methodology to estimate how many infections there must have been to match the data. This highlights a fundamental benefit of of our approach relative to alternative approaches that rely on infection cases, which may be under-counted in the data due to the large number of asymptomatic cases or undetected infections.

\subsection{State mandates}

\begin{sidewaysfigure}[t]
\begin{center}
\subfigure[A state imposes strict or loose policies.]{ 
	\includegraphics[width=0.45\textwidth]{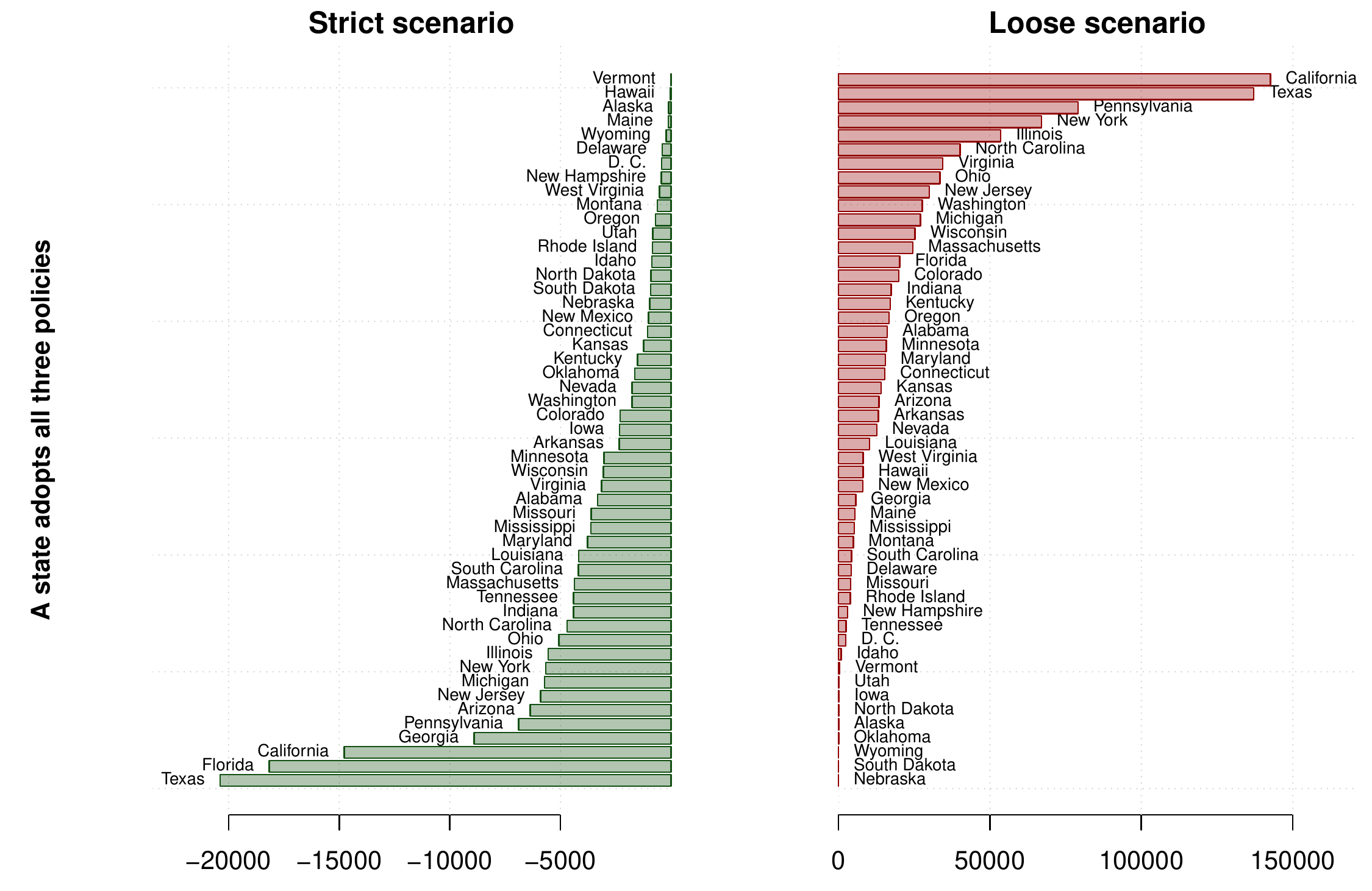}
} \qquad
\subfigure[A state imposes a strict or loose stay-at-home order.]{ 
	\includegraphics[width=0.45\textwidth]{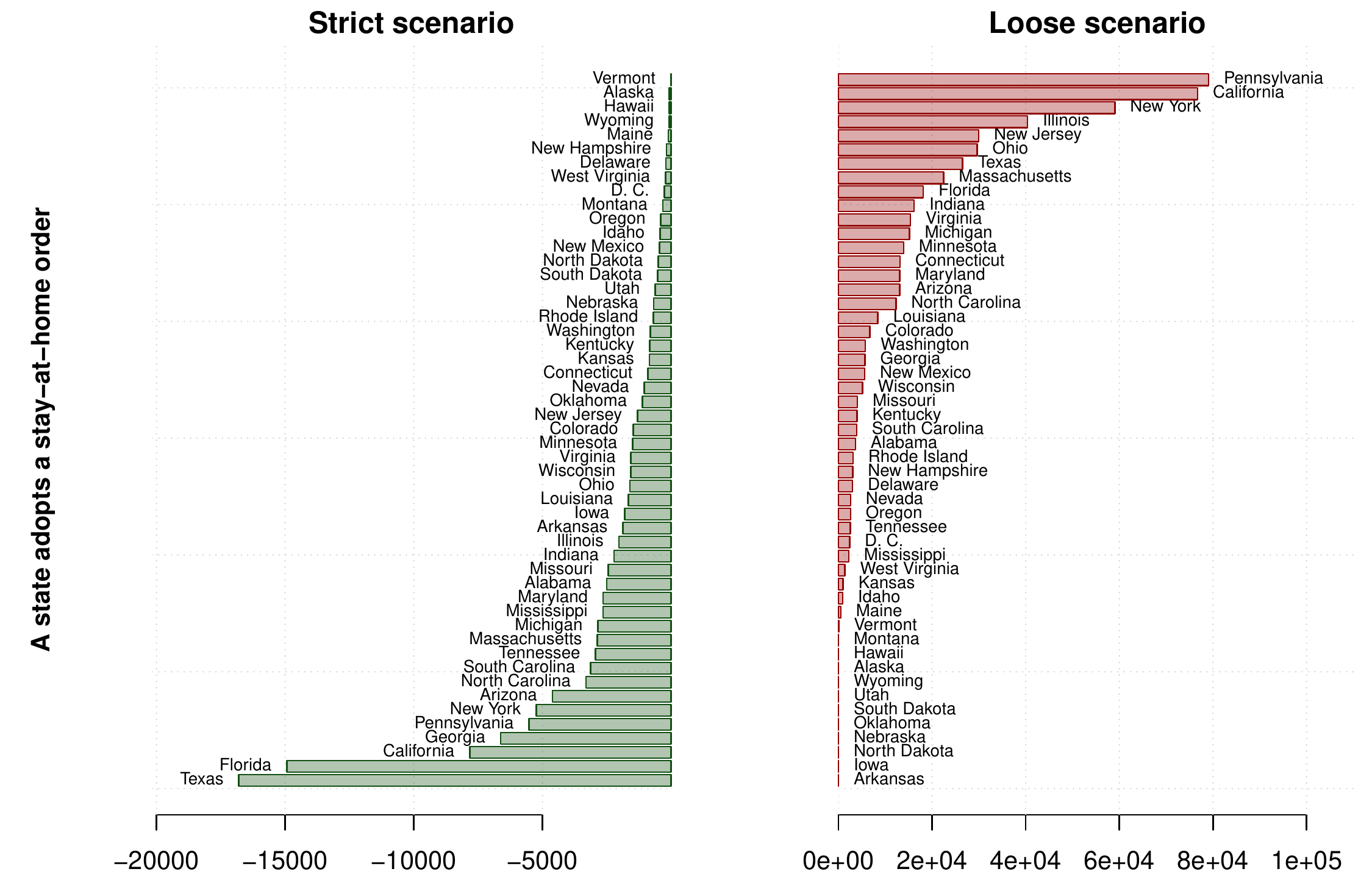}
} \qquad
\subfigure[A state imposes a strict or loose mask mandate.]{ 
	\includegraphics[width=0.45\textwidth]{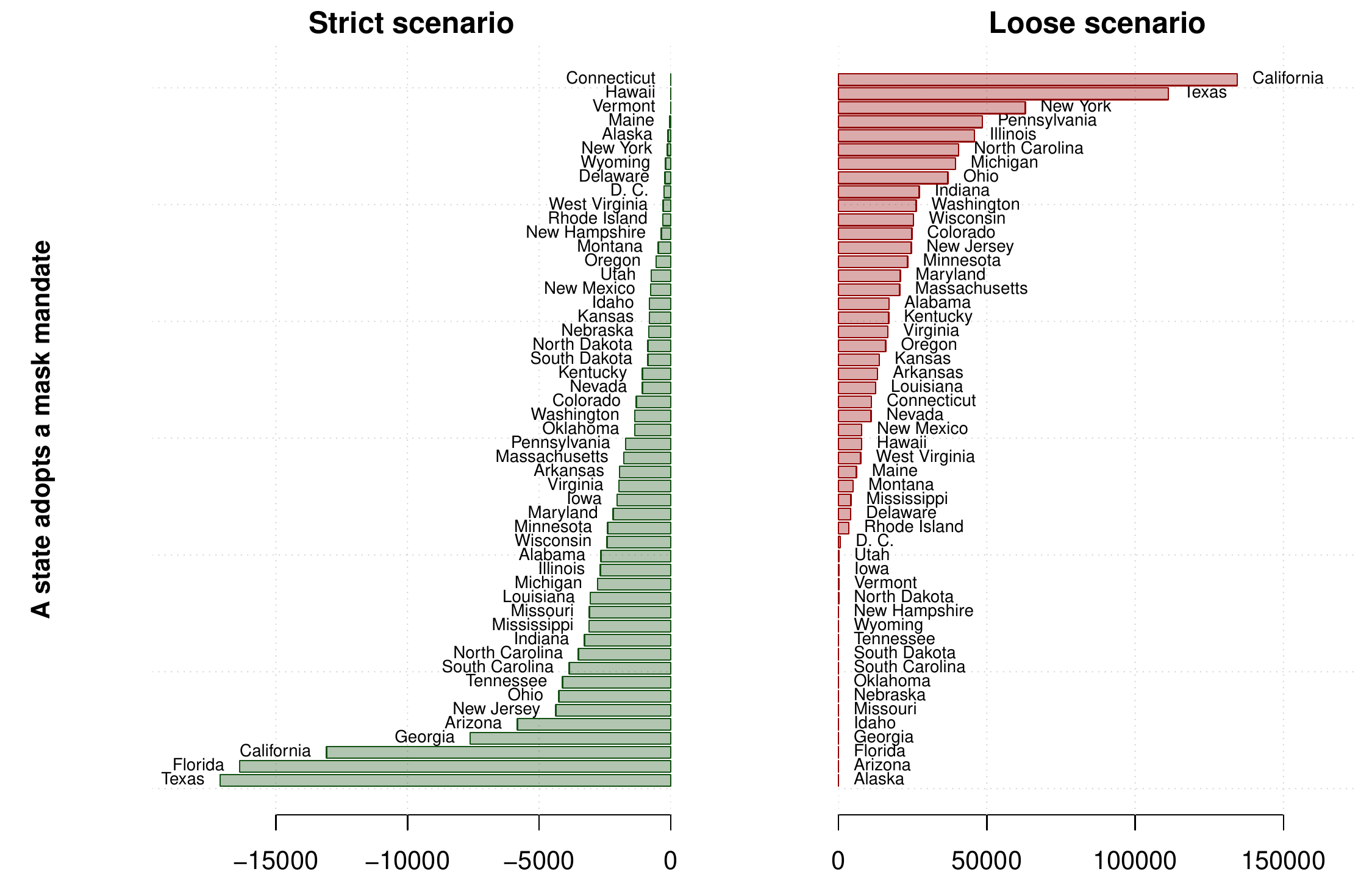}
} \qquad
\subfigure[A state imposes a strict or loose interstate travel ban.]{
	\includegraphics[width=0.45\textwidth]{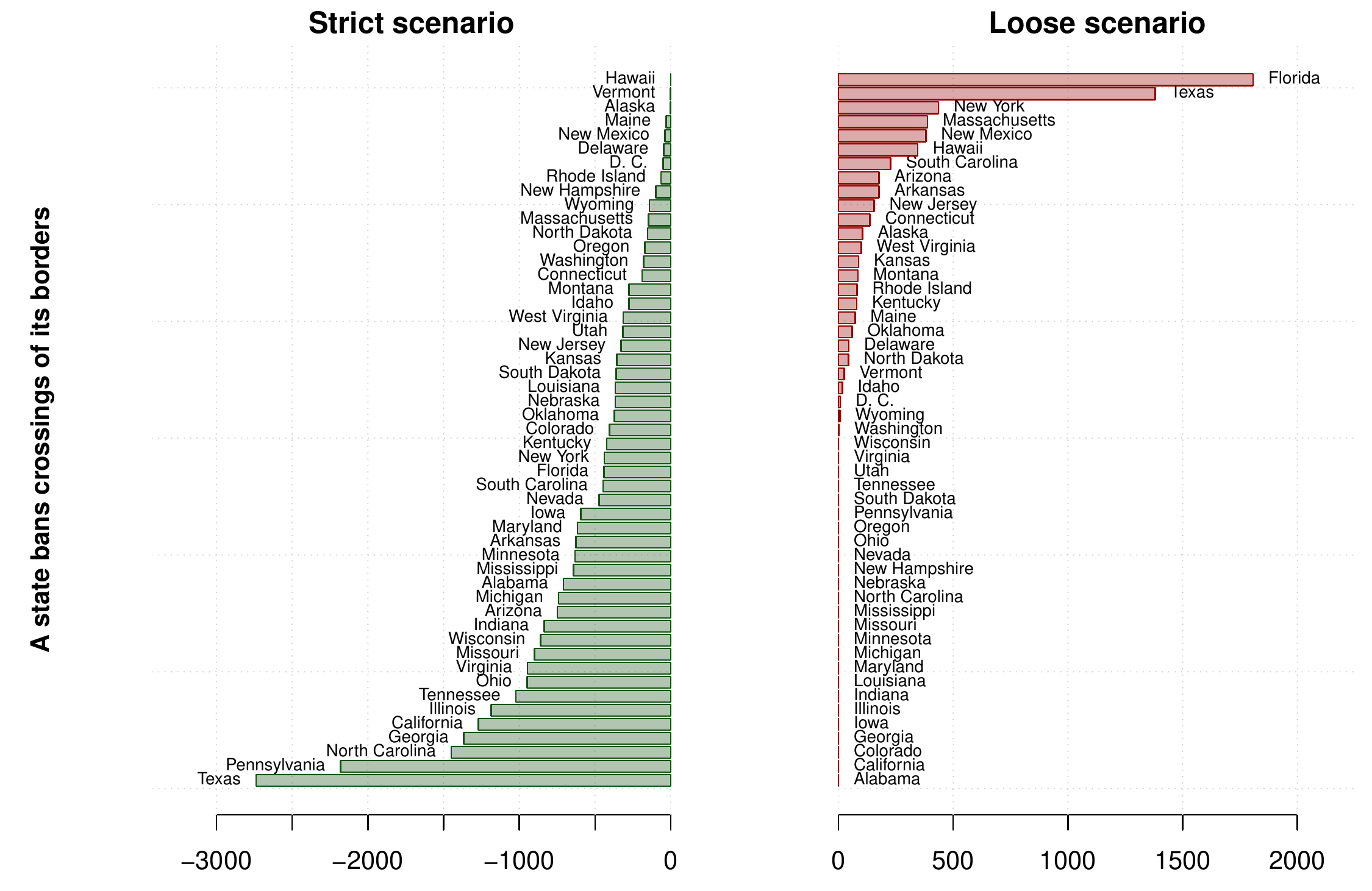}
}
\end{center}
\caption{State-by-state breakdown of excess deaths relative to the baseline in the data in the different counterfactual scenarios in which states deviate individually from their implemented policies. Policies are divided into stay-at-home order (b), mask mandates (c), travel bans (d), and all three policies together (a). Counterfactual policies are divided into a \textit{strict} scenario in which a particular state implements a policy as long as at least one state decides to do so, and a \textit{loose} scenario in which a state does not implement a particular policy. Counterfactual death counts are computed through the methodology detailed in Appendix \ref{app:cf}.} 
\label{fig:state}
\end{sidewaysfigure}

Our counterfactual experiments indicate that the actions taken by individual states benefited both the states that implemented the policies and the U.S. as a nation. We find that the U.S. would have been much worse off if no state had adopted any containment policies. In a hypothetical scenario in which no state had adopted any policies, Table \ref{tbl:fed} indicates that the U.S. would have observed 1,044,270 additional deaths due to COVID-19. Our results show that states that implemented strict containment policies were able to contain the spread of the disease. Figure \ref{fig:federal} shows that a state like New York, which had one of the longest running stay-at-home and mask mandates, would have experienced only 20\% fewer death cases if all states had enforced all three strict policies simultaneously.  Figure \ref{fig:state} also shows that a state like California, which adopted the earliest stay-at-home policy in the country, would have recorded 142,596  additional virus deaths -- more than 7-times the state's end-of-November toll -- if it had not adopted any containment policies at all. These results suggest that the measures adopted by individual states to contain the disease were highly effective and shielded states from inaction from other states.

Figure \ref{fig:state} also shows that states that adopted weak containment policies could have prevented a substantial number of COVID-19 fatalities by imposing stricter policies. Two states in particular stand out: Florida and Texas. These states could have prevented 18,165 deaths (96\% of all death cases in Florida) and 20,385 (95\% of all death cases in Texas), respectively, if they had implemented strict versions of the containment policies we consider in this paper.

It is unrealistic to expect that most states would have enacted strict lockdowns and travel restrictions because these policies carry heavy economic burdens. Mask mandates, on the other hand, offer an effective alternative without significant economic costs.
We find that Texas could have prevented 80\% of its COVID-19 fatalities if the state had adopted a mask mandate as early as Connecticut did on April 17, 2020. 
We find that the twelve states that did not adopt a mask mandate over our sample period could have also significantly benefited from following Connecticut's lead on an early mask mandate. 
Figure \ref{fig:state} suggests that adopting early mask mandates could have prevented as many COVID-19 deaths as 
110 in Alaska (80\% of all COVID-19 deaths in the state), 
5.827 in Arizona (87\%), 
16,373 in Florida (87\%), 
7,616 in Georgia (78\%), 
809 in Idaho (86\%), 
3,096 in Missouri (78\%), 
834 in Nebraska (81\%), 
1,368 in Oklahoma (78\%), 
3,862 in South Carolina (88\%), 
872 in South Dakota (91\%), 
4,108 in Tennessee (88\%), 
and 191 in Wyoming (76\%). 
Even states that otherwise had strict policies in place could have benefitted from adopting earlier mask mandates. Our counterfactuals suggest that California, whose mask mandate went into effect on June 19, could have prevented 13,078 virus deaths (68\% of the state's end-of-November death toll) if it had adopted a mask mandate as early as Connecticut did on April 17.
The effectiveness of state mask mandates is further demonstrated by the dashed red lines of Figure \ref{fig:inf_r0} that show that the United States would have experienced accelerated virus transmission beginning in July 2020, with the national effective reproduction rate consistently above one, if no state had adopted mask mandates.
Our results show that state mask mandates can be highly effective even in the absence of federal policies. 




\subsection{Interstate mobility}

The results of Table \ref{tbl:fed} and Figures \ref{fig:federal}--\ref{fig:state} suggest that interstate travel bans are not very effective in preventing COVID-19 deaths. We find that there are two reasons driving the low effectiveness of interstate travel bans. 

First, interstate travel bans were often imposed late, only once the state regulator became aware of the virus and the virus had already taken hold in the state's population. Table \ref{tbl:fed} shows that 8,149 deaths could have been prevented in the U.S. if all interstate travel had been banned by March 17, 2020. At that point, however, Figure \ref{fig:inf} in Appendix \ref{sec:data} shows that most states must have had several infected individuals so that the virus was already spreading in the states. Banning interstate travel at that point would not have prevented the spread of the virus. 
It may may be different, however, if interstate travel were banned earlier. Early enough to prevent the virus from entering certain states. We run an additional counterfactual in which we ask how many virus deaths could have been prevented if the federal regulator banned all interstate travel on the first day of our sample, February 12. The results, summarized in Table \ref{tbl:fed}, show that an additional 1,338 deaths could have been prevented with an early interstate travel ban. This result indicates that travel bans are more effective if they are implemented early on.

Second, some externalities arise when individuals can move across state lines and these externalities reduce the effectiveness of travel bans. In our model, interstate travel allows for the transfer of infected population across states. 
Consider South Carolina, for example. 
On an average day, our estimates suggest that 96,242 South Carolinians travel or commute out-of-state while 83,572 out-of-state residents travel or commute into South Carolina. As a result, South Carolina is a net exporter of individuals in our model.
Now, Figure \ref{fig:federal} suggests that South Carolina would have recorded 4\% more virus death cases if all interstate travel had been banned on March 17. Why is this the case? We find that this is driven by the fact that, when population is not allowed to cross state borders, infected individuals are forced to stay within the home state. The higher concentration of infected population in those state leads to an accelerated spread of the disease and therefore also to higher infections and death cases. This is showcased in Figure \ref{fig:flows} of Appendix \ref{sec:flows}, in which we plot the estimated cumulative number of infections in South Carolina both under the baseline and in the counterfactual in which a strict federal interstate travel ban is imposed. 

Consider Wyoming, on the other hand. Wyoming is a net importer of travelers and commuters, receiving on an average day a net inflow of around 8,750 out-of-state individuals. Figure \ref{fig:state} indicates that Wyoming would have recorded 56\% fewer death cases if the state had banned inbound travelers and commuters from out-of-state on March 17. Figure \ref{fig:flows} of Appendix \ref{sec:flows} suggests a similar mechanism in this case. Without a travel ban, Wyoming received infected population from other states and this resulted in a higher number of infections in the state.

Our experiments indicate that interstate mobility enables that those states that are net receivers of out-of-state travelers and commuters record higher numbers of death cases than they would in the absence of interstate mobility, while states that are net exporters of travelers and commuters record lower number of virus fatalities.
Policies that restrict interstate mobility do not resolve this externality because they only redistribute infection cases among states. 
Our results suggest that late restrictions of interstate mobility can be counterproductive because they allow for the virus to spread in an accelerated fashion in some states rather than allowing for a balanced distribution of the virus across states. However, we find this is only a minor adverse consequence of policies that restrict interstate mobility. 
In additional unreported experiments, we find that only around 1,700 of the close to 130,000 virus deaths that could have been prevented with a federal stay-at-home order by March 19 are explained by the fact that the order would discourage individuals from commuting and traveling across states.


All in one, our study shows that interstate travel restrictions can be effective for states that normally receive a large number of travelers or commuters, or at the federal level if all interstate travel is banned early enough to prevent the virus from taking hold in different states.

\section{Conclusion}

We show that more than two-thirds of all COVID-19 death cases is the U.S. were preventable had the federal government followed the leads of several states that took early actions to contain the virus. 
Our results indicate that, in the absence of a unified federal approach, the policies enacted in individual states were effective and resulted in reduced virus fatalities.
This benefited both the individual states and the U.S. as a whole. 
As a lesson for state and federal regulators in handling the COVID-19 pandemic going forward, our results show that mask mandates are highly effective policy tools, especially at times when the virus reproduction numbers are large.


\bibliography{biblio}
\bibliographystyle{jmr}

\newpage

\appendix 

\thispagestyle{empty}
\setcounter{section}{0}

\setcounter{page}{1}

\section*{}

\hrule
\vspace{.5cm}

\begin{center}
\begin{large}

{\bf---Online Appendix --}

\end{large}

\bigskip

{\bf \Large Preventing COVID-19 Fatalities:  State versus Federal Policies}

\vspace{.3cm}

Jean-Paul \textsc{Renne}, Guillaume \textsc{Roussellet}, and Gustavo \textsc{Schwenkler}

\vspace{.5cm}
\hrule

\end{center}

\vspace{1cm}

\section{Model}\label{app:Model_Formulation}


\subsection{Single-state model}

We first introduce a single-state model to provide an overview of the assumptions underlying the stochastic evolution of the COVID-19 disease. This provides a basis that we extend in the following section to account for a network of $N$ states.

\subsubsection{From micro assumptions to aggregate dynamic equations}

Consider individual $j$ that was infected on date $t-1$ (i.e., $j \in \{1,\dots,I_{t-1}\}$). We assume that the number of people infected by this individual between dates $t-1$ and $t$ follows a Poisson distribution: $i_{j,t} \sim i.i.d.\, \mathcal{P}\left(\frac{S_{t-1}}{N} \beta_{t-1}\right)$.
Using the fact that a sum of independent Poisson-distributed variables is Poisson, the total number of people infected between dates $t-1$ and $t$ is $i_t \sim i.i.d.\, \mathcal{P}\left(\frac{S_{t-1}}{N}I_{t-1} \beta_{t-1}\right)$.

If $j$ is infected on date $t$, then the probability that she dies between dates $t$ and $t+1$ is $\delta$ (i.i.d. Bernoulli) and, if she does not die, the probability she recovers between dates $t$ and $t+1$ is $\gamma/(1 - \delta)$. (In such a way that the probability she recovers if $\gamma$.)

The cumulated number of dead people on date $t$ ($D_t$) is given by $D_t = D_{t-1} + d_t$,
where $d_t$ is the number of deaths taking place on date $t$. Under the assumptions stated above, $d_t$ follows a binomial distribution: $d_t \sim \mathcal{B}(I_{t-1},\delta)$.
Moreover, the number of people who recover between dates $t-1$ and $t$ is $r_t \sim \mathcal{B}\left(I_{t-1}-d_t,\frac{\gamma}{1-\delta}\right)$.

We have $S_t = S_{t-1} - i_{t}$, $I_t = I_{t-1} + i_t - d_t - r_t$, and $R_t = R_{t-1} + r_t$,
where $S_t$ and $R_t$ respectively denote the number of susceptible and recovered persons as of date $t$. (It is easily checked that $S_t + I_t + R_t + D_t = S_{t-1} + I_{t-1} + R_{t-1} + D_{t-1}$.)

\subsubsection{Transmission rate dynamics}


The transmission rate $\beta_t$ is assumed to follow a non-negative square root process whose Euler discretization reads:
\begin{equation}\label{eq:Beta_dyna}
\beta_t = \mathbb{E}_{t-1}(\beta_t) + \sigma\sqrt{  \Delta t \mathbb{E}_{t-1}(\beta_t)} \varepsilon_{\beta,t},
\end{equation}
where $ \varepsilon_{\beta,t} \sim \mathcal{N}(0,1)$ and $\mathbb{E}_{t-1}(\beta_t) = \beta_{t-1} + \kappa(\beta - \beta_{t-1})$. While the approximate specification (\ref{eq:Beta_dyna}) does not preclude negative $\beta_t$s, the probability associated to such an event remain extremely low, especially at high (daily) frequency. (In practice, at each iteration---i.e. date $t$---of the filtering approaches presented in \ref{app:statespace_singlestate} and \ref{app:statespace}, negative predicted $\beta_t$'s are replaced by zero.)


\subsubsection{State-space model}\label{app:statespace_singlestate}

Let us now write the state-space representation of the model in a context where only $D_t$ is observed. The vector of latent variables is $[\beta_t,S_t,I_t,R_t]'$.

The measurement equation is $\Delta D_t = \delta I_{t-1} + \varepsilon_{D,t}$.
The transition equations are:
\begin{eqnarray}
\left[
\begin{array}{c}
S_t \\
I_t \\
R_t \\
\beta_t 
\end{array}
\right]
&=& 
\left[
\begin{array}{cccc}
1 & 0 & 0 & 0 \\
0 & 1-\delta-\gamma & 0 & 0 \\
0 & \gamma & 1 & 0 \\
0 & 0 & 0 & 1 - \kappa
\end{array}
\right]
\left[
\begin{array}{c}
S_{t-1} \\
I_{t-1} \\
R_{t-1} \\
\beta_{t-1} 
\end{array}
\right] + \nonumber\\
&&\left[
\begin{array}{c}
-\frac{S_{t-1}}{N} \beta_{t-1}I_{t-1}  \\
+\frac{S_{t-1}}{N} \beta_{t-1}I_{t-1}  \\
0 \\
\kappa \beta
\end{array}
\right] +
\left[
\begin{array}{c}
\varepsilon_{S,t}\\
\varepsilon_{I,t}\\
\varepsilon_{R,t}\\
\sigma \sqrt{\beta_{t-1} \Delta t} \varepsilon_{\beta,t}
\end{array}
\right],
\end{eqnarray}
%
%
with $\varepsilon_{\beta,t} \sim \,i.i.d.\,\mathcal{N}(0,1)$ and
\begin{align*}
\left[
\begin{array}{c}
\varepsilon_{D,t}\\
\varepsilon_{S,t}\\
\varepsilon_{I,t}\\
\varepsilon_{R,t}
\end{array}
\right] = 
\left[
\begin{array}{c}
d_t - \delta I_{t-1}\\
- i_t +  \frac{S_{t-1}}{N}I_{t-1} \beta_{t-1}\\
i_t - d_t - r_t  - \frac{S_{t-1}}{N}I_{t-1} \beta_{t-1} + (\delta + \gamma)I_{t-1} \\
r_t -\gamma I_{t-1}\\
\end{array}
\right].
\end{align*}
Appendix\,\ref{App:singleState} shows that $\mathbb{V}ar_{t-1}\left(\left[\varepsilon_{D,t}, \varepsilon_{S,t}, \varepsilon_{I,t}, \varepsilon_{R,t}, \varepsilon_{\beta,t} \right] \right)$ is equal to
\begin{equation}\label{eq:cov_single}
I_{t-1}
\left[
\begin{array}{ccccc}
\delta(1-\delta) & 0 & -\delta(1-\delta-\gamma) & -\delta\gamma & 0\\
0 & \frac{S_{t-1}}{N} \beta_{t-1}& -\frac{S_{t-1}}{N} \beta_{t-1} & 0 & 0\\
-\delta(1-\delta-\gamma)& -\frac{S_{t-1}}{N} \beta_{t-1}& \frac{S_{t-1}}{N} \beta_{t-1}  + \nu  & - \gamma(1 - \gamma-\delta)& 0\\
 -\delta\gamma & 0 & - \gamma(1 - \gamma-\delta) & \gamma - \gamma^2& 0\\
 0 & 0 & 0 & 0 & 1
\end{array}
\right],
\end{equation}
where $\nu = \frac{(1 -\delta-\gamma)^2\delta+\gamma(1-\delta-\gamma)}{1-\delta}$.

\subsection{Multi-state model}

We now consider a $N$-state model. Inhabitants may travel across states for commuting or other reasons -- we refer to the latter as ``traveling.'' In terms of notations, superscript $j$ refers to a given state. Variables without superscripts denote $N$-dimensional vectors. We denote by ${\bf p}=[p_1,\dots,p_N]'$ the vector of the state population sizes. $S_t$,  $I_t$,  $R_t$ and  $D_t$ are $N$-dimensional vectors gathering the number of susceptible, infected, recovered, and deceased people in each state.

\subsubsection{Policy restrictions}\label{sub:restrictions}


The transmission rates $\beta_{j,t}$ and the flow probabilities are impacted by the containment policies implemented in the different states. We consider three containment policies: mask mandates, stay-at-home orders, and travel bans. The effect of each policy is captured through the binary variables $\theta_{t,M}^j$, $\theta_{t,S}^j$ and $\theta_{t,T}^j$, respectively valued in $\{\theta_{M}^{low},1\}$, $\{\theta_{S}^{low},1\}$, and $\{\theta_{T}^{low},1\}$. The parameters $\theta_{M}^{low}$, $\theta_{S}^{low}$, and $\theta_{T}^{low}$ are strictly lower than one; they reflect the effects of the containment policies. More precisely:
\begin{itemize}
	\item The transmission rate $\beta_{j,t}$ is reduced when mask mandates or stay-at-home policies are implemented. Formally, it is of the form $\beta^0_{j,t}\theta_{t,M}^j\theta_{t,S}^j$, where $\beta^0_{j,t}$ is an exogenous transmission rate following the dynamics depicted by \eqref{eq:Beta_dyna}. Given that the $\theta$ variables are equal to one when the policies are not in place, it follows that $\beta^0_{j,t}$ coincides with the effective transmission rate ($\beta_{j,t}$) when mask mandates and stay-at-home policies are not implemented.
	\item The probability that a given inhabitant of State $j$ commutes to State $k$, that is $w^{com}_{j,k,t}$, is of the form $w^{com}_{j,k}\theta^k_{S,t}$. Similarly, the travel probability is given by $w^{trav}_{j,k,t} = w^{trav}_{j,k}\theta^k_{S,t}\theta^k_{T,t}$. These probabilities are therefore lower when \textit{(i)} stay-at-home orders are in place or \textit{(ii)} when travel bans are enforced in the visited state. 
\end{itemize}

\subsubsection{Traveler flows} \label{app:trav_flows}

	The variables $\mbox{Flow}_{trav,S,t}^j$, $\mbox{Flow}_{trav,I,t}^j$, and $\mbox{Flow}_{trav,R,t}^j$ are net travel inflows of susceptible, infected, and recovered populations, respectively. 
		
We denote by $w_{trav,t}^{k,j}$ the average fractions of the date-$t$ population of State $k$ that travels to State $j$; ${ \bf e}_j$ is the $j^{th}$ column of the $N \times N$ identity matrix; ${\bf 1}$ is a $N \times 1$ vector of ones; $w_{trav,t}^{j,\bullet}$ and $w_{trav,t}^{\bullet,j}$ respectively denote the $j^{th}$ row vector and column vector of $W_{trav,t}$.
	Consistent with the assumptions made in \ref{sub:restrictions}, we have:
	\begin{equation}\label{eq:Wtravel}
	W_{trav,t} = \tau_{trav} W_{trav} \,\mbox{\bf d}(\theta_{T,t}) \,\mbox{\bf d}(\theta_{S,t}).
	\end{equation}
	Here, $\tau_{trav}$ measures the average number of days that a traveler spends in the visited states.
	Using the Poisson approximation of the binomial distribution, we consider that the number of outward travelers is drawn from Poisson distributions. For example, the number of susceptible individuals traveling from State $k$ to State $j$ between dates $t$ and $t+1$ is:
	\begin{equation}\label{eq:Fskj}
	\mbox{Flow}_{trav,S,t}^{k,j} \sim i.i.d. \, \mathcal{P}(w_{trav,t}^{k,j}S_{t}^{k}).
	\end{equation}
	This implies in particular that the net number of inhabitants traveling into State $j$ between dates $t$ and $t+1$ (denoted by $\mbox{Flow}_{trav,S,t}^j$) is such that:
	\begin{eqnarray*}
	\phi_{trav,S,t}^j := \mathbb{E}(\mbox{Flow}_{trav,S,t}^j|S_{t}) &=& \left(\sum_{k \ne j}  w_{trav,t}^{k,j} S_{t}^{k}\right) - \left(\sum_{j \ne k} w_{trav,t}^{j,k}\right) S_{t}^{j}\\
	&=& \left(w_{trav,t}^{\bullet,j} - ({w_{trav,t}^{j,\bullet}} { \bf 1}){\bf e}_j\right)'S_{t}.
	\end{eqnarray*}
	Using the convention $w_{trav,t}^{j,j}=0$, it follows that the $N$-dimensional vector $\phi_{trav,S,t}$ is given by $\phi_{trav,S,t} = \Omega_{trav,t}S_t$, where
	\begin{equation}\label{eq:Omegatrav}
	\Omega_{trav,t} = W_{trav,t}' - \mbox{\bf d}(W_{trav,t} {\bf 1}).
	\end{equation}
	By the same token, and with obvious notations for $\phi_{trav,I,t}$ and $\phi_{trav,R,t}$:
	$\phi_{trav,I,t} = \Omega_{trav,t} I_t$ and $\phi_{trav,R,t} = \Omega_{trav,t} R_t$.

\subsubsection{Commuter flows} \label{app:comm_flows}

	Interstate commuters are people who spend a fraction $\tau$ of each day in another state. Consider the infected inhabitants of State $k$ working in State $j$. They contaminate less people in State $k$ because they spend less time in that state. But they may also contaminate people in State $j$ because  they spend some time in that state while commuting. We respectively denote by $\mbox{Flow}_{com,S,t}^{j \leftarrow}$,  $\mbox{Flow}_{com,I,t}^{j \leftarrow}$, and $\mbox{Flow}_{com,R,t}^{j \leftarrow}$ the commuting inflows of susceptible, infected and recovered people in State $j$. Outflows are given by $\mbox{Flow}_{com,S,t}^{j \rightarrow}$,  $\mbox{Flow}_{com,I,t}^{j \rightarrow}$ and $\mbox{Flow}_{com,R,t}^{j \rightarrow}$.

	Let us denote by $W_{com,t}$ the ``commute'' matrix; that is, the matrix whose component $(i,j)$, denoted by $w_{com,t}^{i,j}$, is the fraction of the date-$t$ population of State $k$ that commutes to State $j$. On date $t$, the number of susceptible people commuting from State $k$ to State $j$ is:
	\begin{equation}\label{eq:FSCkj}
	\mbox{Flow}_{com,S,t}^{k,j} \sim i.i.d. \, \mathcal{P}(w_{com,t}^{k,j}S_{t}^{k}).
	\end{equation}
	This implies in particular, with obvious vectorial notations, that:
	\begin{equation}\label{eq:PhiSCleft}
	\phi_{com,S,t}^{\leftarrow} := \mathbb{E}(\mbox{Flow}^{\leftarrow}_{com,S,t}|S_t)  = W_{com,t}'S_{t},
	\end{equation}
	with (consistently with the assumptions made in \ref{sub:restrictions}):
	\begin{equation}\label{eq:Wtravel}
	W_{com,t} =  W_{com} \,\mbox{\bf d}(\theta_{S,t}).
	\end{equation}
	where $W_{com}$ is the commute matrix that would prevail under no containment policies.
	
	By the same token:
	\begin{equation}\label{eq:PhiSCright}
	\phi_{com,S,t}^{\rightarrow}  := \mathbb{E}(\mbox{Flow}^{\rightarrow}_{com,S,t}|S_t)  = \mbox{\bf d}(W_{com,t}{\bf 1})S_{t}.
	\end{equation}	
	
	All in all, if we denote by $\mbox{Flow}_{com,S,t}$ the vector of time-weighted commuters net inflows, we have:
	\begin{equation}\label{eq:FlowSC}
	\mathbb{E}(\mbox{Flow}_{com,S,t}|S_t) = \Omega_{com,t} S_t,
	\end{equation}
	with
	\begin{equation}\label{eq:OmegaC}
	\Omega_{com,t} = \tau_{com} W_{com,t}' - \tau_{com} \mbox{\bf d}(W_{com,t}{\bf 1}).
	\end{equation}

\subsubsection{Transmission rate dynamics}


Each state $j$ features an autonomous $\beta^0_{j,t}$ process (see \ref{sub:restrictions}). These variables, gathered in vector $\beta^0_t$, follow non-negative square-root processes whose dynamics is approximated by:
$$
\beta^0_{j,t} \approx \beta^0_{j,t-1} + \kappa (\beta - \beta^0_{j,t-1})\Delta t + \sigma \sqrt{\Delta t \beta^0_{j,t-1} } \varepsilon^j_{t},
$$
with $\varepsilon_{t} \sim \,i.i.d.\,\mathcal{N}(0,\Sigma)$, where the diagonal elements of $\Sigma$ are ones and the extra-diagonal entries are set to $\rho$.

\subsection{Effective reproduction number}\label{App:EffR0}

The effective reproduction number is defined as the average number of persons contaminated by an infected person. It is computed as:
\begin{equation}\label{eq:EffectiveRt}
\mathcal{R}_{j,t} = \theta_{t,M}^j\theta_{t,S}^j \underbrace{\frac{\beta^0_{j,t}}{\delta + \gamma} }_{= \mathcal{R}_{0,t}}  \frac{S_{j,t}}{p_j},
\end{equation}
where $\mathcal{R}_{0,t}$ is usually called basic reproduction number. The latter, in turn, is defined as the average number of people the first infected individual of a population infects when (i) no containment policies are in place and (ii) all individual (except the first case) is susceptible.

	\subsubsection{State-space model}\label{app:statespace}
	
	The state-space model is characterized by 
	$$
\mathbb{E}_{t-1}\left(\left[
\begin{array}{c}
D_{t}\\
S_{t}\\
I_{t}\\
R_{t}\\
\beta_t^0
\end{array}
\right]\right)\quad \mbox{and}\quad
\mathbb{V}ar_{t-1}\left(\left[
\begin{array}{c}
D_{t}\\
S_{t}\\
I_{t}\\
R_{t}\\
\beta_t^0
\end{array}
\right]\right).
$$
We have:
\begin{eqnarray}
&&\mathbb{E}_{t-1}\left(
\left[
\begin{array}{c}
D_{t}\\
S_{t}\\
I_{t}\\
R_{t}\\
\beta_t^0
\end{array}
\right]
\right)= 	\left[
\begin{array}{c}
0\\
0\\
0\\
0\\
\kappa\beta {\bf 1}
\end{array}
\right]+  \left[
	\begin{array}{ccccc}
	{\bf Id} & 0 & \delta {\bf Id} & 0 & 0\\
	0 & {\bf Id} & 0 & 0 & 0 \\
	0 & 0 & (1-\delta -\gamma){\bf Id}  & 0 & 0 \\
	0 & 0 & \gamma {\bf Id} & {\bf Id} & 0 \\
	0 & 0 & 0 & 0 & (1 - \kappa){\bf Id}
	\end{array}
	\right]
\left[
\begin{array}{c}
D_{t-1}\\
S_{t-1}\\
I_{t-1}\\
R_{t-1}\\
\beta_{t-1}^0
\end{array}
\right]
	+ \nonumber\\
	&&
	\left[
	\begin{array}{c}
	0 \\
	-{\bf Id}\\
	+{\bf Id}\\
	0\\
	0
	\end{array}
	\right] \left(\theta_{S,t-1}\odot \theta_{M,t-1}  \odot \beta_{t-1}^0 \odot \frac{\bf 1}{\bf p} \odot ([{\bf Id} + \Omega_{t-1}] I_{t-1} )\odot ([{\bf Id} + \Omega_{t-1}] S_{t-1} )  \right),
\end{eqnarray}
where $\Omega_{t-1}=\Omega_{com,t-1}+\Omega_{trav,t-1}$, where the latter two matrices are respectively defined in equations\,(\ref{eq:Omegatrav}) and (\ref{eq:OmegaC}). Notice that the state-space ends up being of size $5\times N$, where $N$ is the number of states, so 255 in our application. 

Appendix \ref{App:multiState} details the computation of the conditional variance of $[D_{t},S_{t},I_{t},R_{t},\beta_t^0]$ (see equation\,\ref{eq:VarSvector} in \ref{app:subCondiVar}).

The previous equations constitute the set of transition equations of the state-space. We complete the formulation with the measurement equations being only the (seasonally adjusted) time series of fatalities per day, for each state, which we denote by $D_t^{obs}$. We assume that the number of deaths per day is measured nearly perfectly, such that:
\begin{equation}
	D_t^{obs} = D_t + \eta_t, \quad \mbox{where} \quad \eta_t\sim \mathcal{N}(0, 0.001^2\mathrm{\mathbf{Id}})\,.
\end{equation}
Since the state-space is non-linear, we resort to the extended Kalman filter for estimation. This requires the computation of the Jacobian matrix of $\mathbb{E}_{t-1}[D_{t},S_{t},I_{t},R_{t},\beta_t^0]$ with respect to $[D_{t-1},S_{t-1},I_{t-1},R_{t-1},\beta_{t-1}^0]$, which is closed-form and detailed in Appendix \ref{app:Jacobian_filter}. We then apply a fixed-interval Rauch-Tung-Striebel smoother (backward filter) with the estimated trajectories produced by the filter.

\section{Data \& estimates} \label{sec:data}

Our model features several parameters that we need to fix: The death rate $\delta$, the recovery rate $\gamma$, the parameters $\beta$, $\kappa$, $\sigma$, and $\rho$ governing the dynamics of transmission rates in the states, the effects $\theta_M^{low}$, $\theta_S^{low}$, and $\theta_T^{low}$ of the different containment policies, and the average traveling and commuting flows across states. We proceed as follows to select parameter values. We summarize our parameter estimates in Table \ref{tbl:est}.

We follow \ci{Fernandez-Villaverde_Jones_2020}, \cite{PerezSaez_et_al_2020}, and \cite{Stringhini_et_al_2020} and assume that it takes on average 14 days for an infection to resolve. We assume that after this period, an infected person either recovers and becomes immune, or dies with a probability of 0.6\%. These assumptions imply that $\gamma = \frac{1}{14}$ and $\delta = \frac{0.06\%}{14} = 0.0004$.

\begin{sidewaysfigure}[t]
\begin{center}
\subfigure[Interstate travel network. The size of a node is proportional to the percentage of a state that commutes out-of-state.]{ \label{fig:travel}
	\includegraphics[width=0.45\textwidth]{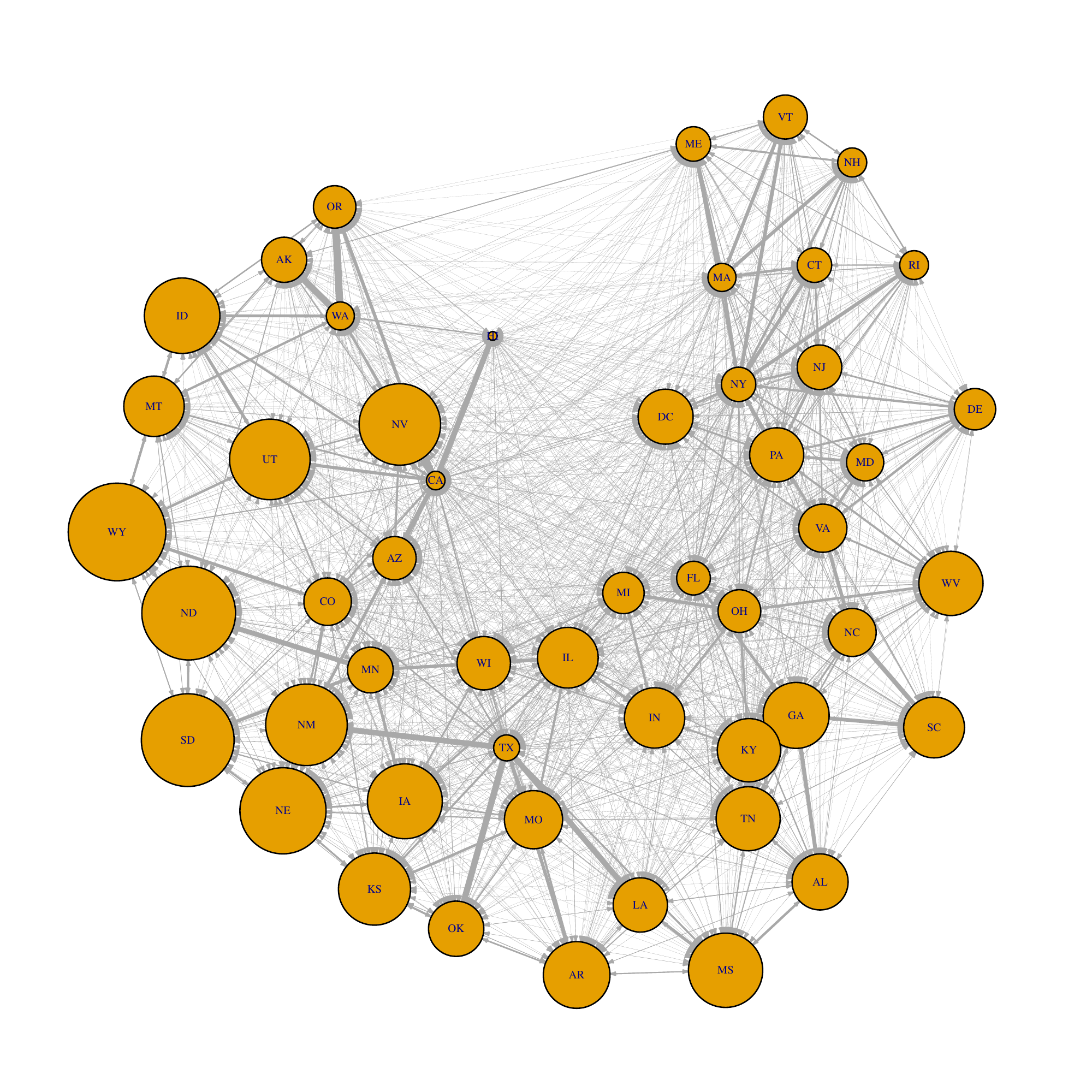}
} \qquad
\subfigure[Interstate commuting network. The size of a node is proportional to the logarithm of the percentage of a state that commutes out-of-state.]{ \label{fig:commute}
	\includegraphics[width=0.45\textwidth]{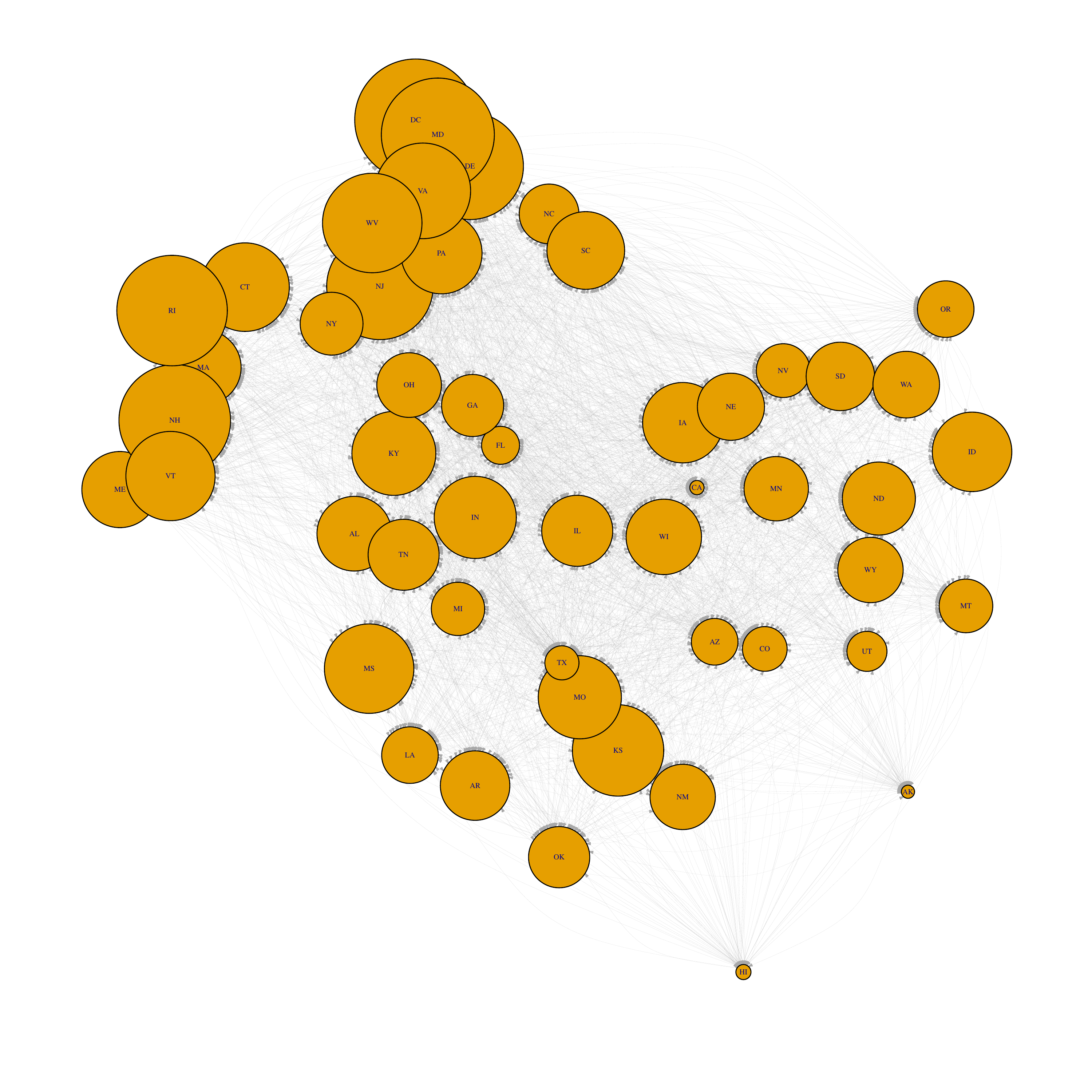}
}
\end{center}
\caption{Estimated interstate travel and commuting flows. For both networks, the width of a link is proportional to the percentage of a state's travel or commuting that takes place between the linked states.}
\end{sidewaysfigure}

We estimate average interstate travel flows from travel and mobility data in the United States.
We collect data on interstate travels from the Traveler Analysis Framework published by the Federal Highway Administration (\url{https://www.fhwa.dot.gov/policyinformation/analysisframework/01.cfm}). We also collect data on state-level mobility and staying-at-home from the Trips by Distance database of the Bureau of Transportation Statistics (\url{https://www.bts.gov/distribution-trips-distance-national-state-and-county-level}). We combine these two databases to compute the percentage of a state's population that stays home before and during the pandemic, as well as the percentage of a state's population that traveled across state boundaries before and during the pandemic. We use these data to determine the travel matrix $W_{trav}$ of Appendix \ref{app:trav_flows}. We illustrate the estimated interstate travel network in Figure \ref{fig:travel}. The size of a node is proportional to the percentage of a state's population that travels outwards and the width of a link is proportional to the percentage of a state's population that travel to the linked state. We assume that an average traveler spends 4 days on vacation. This implies that $\tau_{trav} = 4$.

We measure commuting flows from the 2011-2015 5-Year ACS Commuting Flows table of the U.S. Census (\url{https://www.census.gov/data/tables/2015/demo/metro-micro/commuting-flows-2015.html}). We use these data to estimate the commuter matrix $W_{com}$ of Appendix \ref{app:comm_flows}.
Figure \ref{fig:commute} shows the implied commuting travel network. 
In this figure, however, the size of a node is proportional to the logarithm of the percentage of a state's population that commutes outwards. We assume that out-of-state commuters spend 8 hours each business day and no time during a weekend in the visited state. We also assume that commuters sleep 8 hours a day and during that time infections are not possible. As a result, we set $\tau_{trav} = 0.36 \approx \frac{8 \times 5}{16 \times 7}$.

\begin{sidewaysfigure}[htbp]
\centering
\includegraphics[width = 0.95\textwidth]{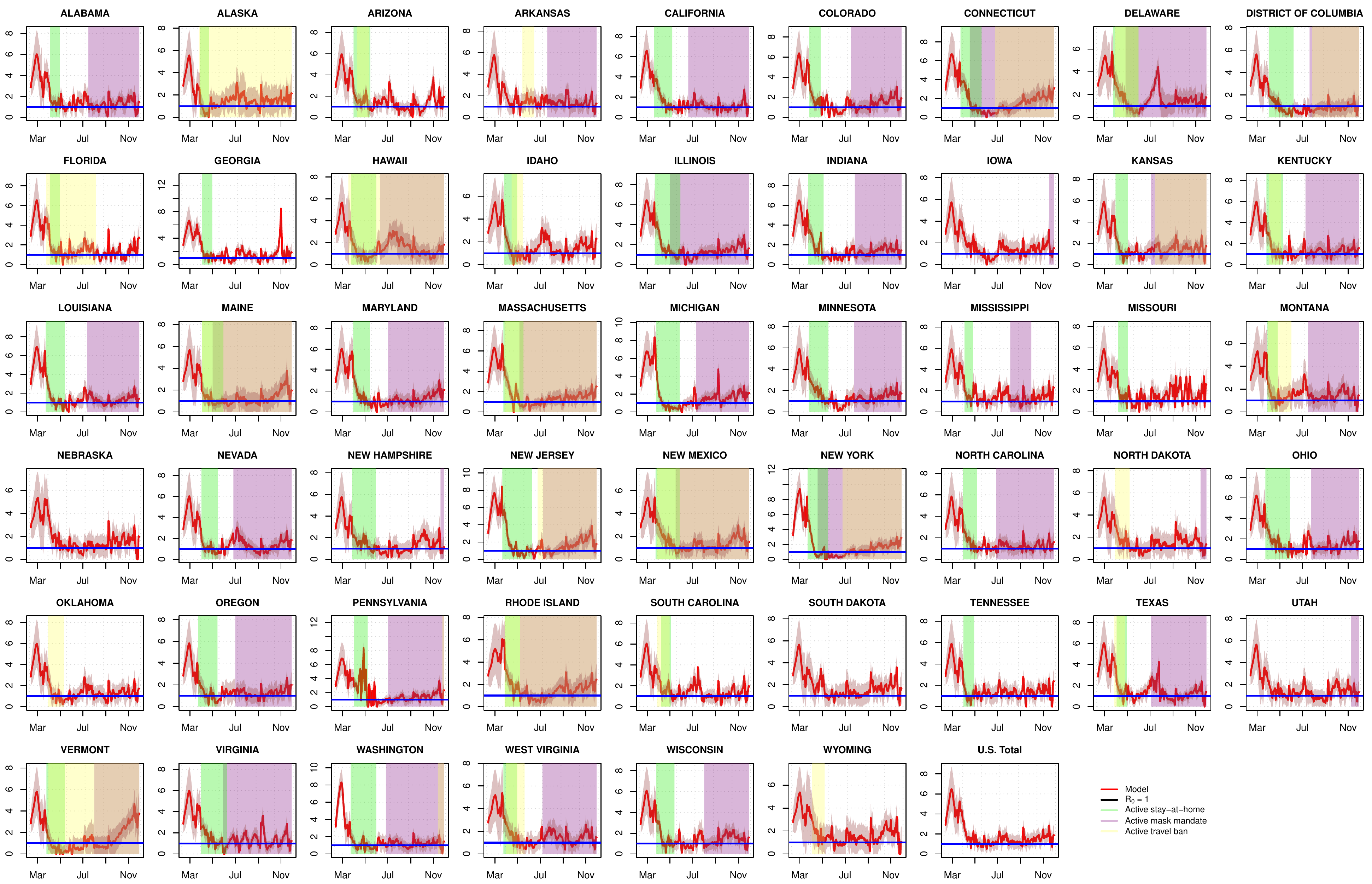}
\caption{Smoothed means of the model-implied effective reproduction numbers $\mathcal{R}_t$ of COVID-19 in U.S. states (see Appendix \ref{App:EffR0} for details). The shaded red areas denote 2-standard-deviation confidence bands. Estimation is performed through forward/backward extended Kalman filtering, using the time series of death counts per state from February 12 to November 30, 2020. Our estimation methodology is detailed in Appendix \ref{sec:data}. The effective reproduction rate estimates correspond to the ratios of estimated $\beta_t^0$ multiplied by the policy dummies and divided by the sum of the daily fatality and recovery rates $(\gamma+\delta)$, scaled by the fraction of a state's population that remains susceptible.
The figure also shows the periods of time in which the different containment policies were active. Green shaded areas correspond to active stay-at-home policies, purple areas to active mask mandates, and yellow areas to active travel bans. Horizontal blue lines correspond the standard value of $\mathcal{R}_t=1$, below which the virus does not reproduce itself explosively.}
\label{fig:R0}
\end{sidewaysfigure}

We calibrate the parameters $\theta_M^{low}$, $\theta_S^{low}$, and $\theta_T^{low}$ to match mobility and mask usage data from the United States. We collect data on when the different policies were active in the different states from the National Academy for State Health Policy (\url{https://www.nashp.org/governors-prioritize-health-for-all/}) and the Steptoe COVID-19 State Regulatory Tracker (\url{https://www.steptoe.com/en/news-publications/covid-19-state-regulatory-tracker.html}). Figure \ref{fig:R0} showcases the time periods in which policies were active in the different states.
We also collect data on average mask adoption across U.S. states from the Institute of Health Metrics and Evaluation (\url{https://covid19.healthdata.org/united-states-of-america}).
We assume that a stay-at-home policy is in place for the time period that covers any order for staying at home, sheltering at home, or being safer at home issued by a state governor. We neglect any stay-at-home advisories that are not strictly enforced by law officials. We estimate $\theta_S^{low} = 0.64$ as the reduction factor in the number of short trips taken during the pandemic versus  before the pandemic.\footnote{More precisely, for each state we evaluate the average number of daily trips of less than 10 miles taken in 2019 and compare that number to the average number of daily trips of less than 10 miles taken during the time frame March 15 through April 30, 2020. We compute $\theta_S^{low}$ as the average reduction factor across states.}
We consider a travel ban to be active when a state requires inbound travelers to self-quarantine for an extended period of time. Travel bans are active in our model only if they apply for all states. That is, we neglect any travel ban that only applies for travelers from selected states. We estimate $\theta_T^{low} = 0.10$ as the reduction factor in the number of long trips taken during the pandemic versus  before the pandemic.\footnote{We evaluate the average number of daily trips of more than 100 miles taken in 2019 and compare that number to the average number of daily trips of more than 100 miles taken during the time frame March 15 through April 30, 2020. We compute $\theta_T^{low}$ as the average reduction factor across states.}
Finally, we assume that a mask mandate is active if a state requires the use of masks indoors in public places. We do not consider mask mandates to be active if mask wearing is only recommended or only required outdoors. We estimate $\theta_M^{low} = 0.58$ by assuming that only half of the population adopts mask usage (as suggested by the Institute of Health Metrics and Evaluation) and that an average mask used in the U.S. reduces COVID-19 transmission by 85\% (as suggested by \ci{fischer-et-al-2020}).

We develop a quasi maximum likelihood methodology to estimate the parameters governing the dynamics of the transmission rates $\beta_{j,t}$ from daily data on state-level deaths in the United States for the time period between February 12 through November 30, 2020. 
We remove weekly seasonality patterns observed in COVID-19 fatality records using an STL approach.
Our methodology assumes that daily death counts at the state-level are measures with small measurement errors. 
We take into account all state-level containment policies that were observed over the sample period. 
We write a non-linear state-space representation of the model, gathering $S_t^j$, $I_t^j$, $R_t^j$, $D_t^j$, and $\beta_{j,t}$ for all states simultaneously (255 variables). Filtering is easily performed through the first order extended Kalman filter algorithm.
While we could estimate the speed of reversion $\kappa$ with our methodology, we find that the data prefers to set $\kappa$ arbitrarily close to zero, implying an extremely high persistence for the $\beta_{j,t}$. This results in numerical instabilities. To avoid these issues, we fix $\kappa = 0.001$ so that the first-order autocorrelation of the $\beta_{j,t}$ is $0.999$. We then estimate the remaining parameters $\beta$, $\sigma$, and $\rho$ using our quasi maximum likelihood methodology. The estimates are provided in Table \ref{tbl:est}. We find that our parameter estimates are not very sensitive to alternative choices for the value of $\kappa$.

Figure \ref{fig:death} shows the data-implied death counts in each state as well as the smoothed model-implied death counts. We see that our model performs well at fitting the state-level data. The estimated measurement errors are fairly small. Figures \ref{fig:R0} and \ref{fig:inf} show the smoothed model-implied effective $\mathcal{R}_t$ and cumulated infections for each state. The data pushes our model to showcase high $\mathcal{R}_t$ in the different states. 
The highest $\mathcal{R}_t$ were observed in New Jersey, New York, and Washington in February and March, reaching levels of more than 8.\footnote{Our estimates of the effective reproduction numbers are one-and-a-half to two-times larger than prevailing estimates in \ci{Fernandez-Villaverde_Jones_2020}. We find that this is driven by a key difference in our models. \ci{Fernandez-Villaverde_Jones_2020} assume that, while it take two weeks for an infection to resolve, an infected individual is only contagious for the first 5 days of an infection. We, instead, assume that an infected individual is contagious during the whole infection period. We run several experiments in Appendix \ref{app:sens} in which we study whether are results are sensitive to this assumption, and we find that this is not the case.}
Indeed, we find that the state-level $\mathcal{R}_t$ must have been substantially higher than $1$ for most states through April 2020. Several states, like Delaware, Hawaii, Idaho, Montana, Oklahoma, South Carolina,  Texas, and West Virginia, experienced significant upticks in their Coronavirus $\mathcal{R}_t$ in the summer. 
There have also been upticks in the $\mathcal{R}_t$ in September in states like Arkansas, Florida, Kansas, Michigan, Missouri, North Dakota, Rhode Island, South Dakota, and Virginia.
Some states have been successful at maintaining their $\mathcal{R}_t$ values consistently at or below $1$. The list of states that have accomplished this task include California, D. C., Illinois, Indiana, Maine, Maryland, Massachusetts, Minnesota, New Hampshire, New Mexico, New York, Oregon, and Vermont.
Overall, our findings suggest that the virus spread drastically  in the U.S. through the Fall of 2020. In fitting the data, our model estimates that the number of infections in the different states must have been significantly higher than recorded in the data. These observations suggest that there must have been many undiagnosed infections that facilitated the spread of the disease throughout our sample period. Note that we never use infections data for the estimation or calibration of our model parameters.


\section{Counterfactual experiments} \label{app:cf}

	This section details how our counterfactual experiments are conducted.

	\subsection{Baseline scenario}
	
	The estimated dynamics constitutes our baseline scenario. All of our results are presented as a difference with respect to this baseline scenario. We compare the outcomes in terms of fatalities.
	
	\subsection{``Strict'' and ``loose'' counterfactuals}
	
	\noindent We consider two types of experiments, which we call strict and loose, respectively:
	\begin{itemize}
		\item[$\bullet$] In the strict scenario, we assume that the states start adopting the same policy as that implemented by the earliest state during the sample period and relax it when the latest state does so. The scenarios are as follows:
	\begin{itemize}
		\item \textbf{Stay-at-Home}: Start on March 19 (the day the first stay-at-home order in a state went into effect in California) and end on June 9, 2020 (the day the last stay-at-home order in a state ended in Virginia).
		\item \textbf{Travel ban}: Start on March 17, 2020 (the day the first interstate travel ban in the U.S. went into effect in Hawaii) and keep it active through November 30, 2020 (the last day in our sample when several states had travel restrictions in place).
		\item \textbf{Mask mandate}: Start on April 17, 2020 (the day the first mask mandate in the U.S. went into effect in Connecticut) and keep it active through November 30, 2020 (the last day in our sample when several states had mask mandates in place).
	\end{itemize}	 
	\item[$\bullet$] For the loose scenario, we assume that states do not implement a specific policy at all. We then re-propagate the model with these counterfactual policies according to the methodology described below.
	\end{itemize}
	
	\noindent We conduct these two types of experiments one policy at a time, and one last time all together. 

	\subsection{Joint and state-by-state counterfactuals}
	
	Our analysis is split into joint and state-by-state experiments. For the former, we assume that the federal government imposes on all states the counterfactual policy (strict federal mandate), i.e. to be as strict as the strictest or to do nothing (loose federal mandate). For the latter, we take each state one at a time and assume that only this state follows the strict or loose scenario and compute the counterfactual outcomes one state at a time.
	
	\subsection{Computation of counterfactual scenarios}

Recall that the model parameters and the latent variables ($S_t$, $I_t$, $R_t$ and $\beta^0_t$) are estimated by employing the extended Kalman filter. The observed variables are the numbers of deaths $D_t^{obs.}$, as well as the observed implemented policies $\theta_{obs.}$.

In order to derive our counterfactual outcomes, we also extend our filtering approach. The broad idea is the following: We augment the state vector used at the estimation step -- i.e. $X_t = [D_t,S_t,I_t,R_t,\beta_t^0]$ -- with a similar state vector for a fictitious country ($fict.$), with the same number of states, where the counterfactual policies would be implemented. Critically, we assume that the basic, standardized, shocks affecting the two countries are almost perfectly correlated, implying in particular that the $\beta^0_t$'s are the same for the baseline and fictitious countries.

Specifically, let us denote by $X^*_t$ the state vector corresponding to the fictitious country. The transition equation of the augmented state-space model are:
\begin{equation}\label{eq:transitCounter}
\left[
\begin{array}{c}
X_t\\
X^*_t
\end{array}
\right] = \left[
\begin{array}{c}
\mathcal{E}(X_{t-1},\theta_{obs.})\\
\mathcal{E}(X^*_{t-1},\theta_{fict.})\\
\end{array}
\right] + 
\left[
\begin{array}{c}
\mathcal{V}^{1/2}(S_{t-1},I_{t-1},\theta_{obs.}) \varepsilon_{True,t}\\
\mathcal{V}^{1/2}(S^*_{t-1},I^*_{t-1},\theta_{fict.}) \varepsilon_{fict.,t}
\end{array}
\right],
\end{equation}
where $\theta_{obs.}$ and $\theta_{fict.}$ contain the full trajectories of observed and fictitious policies, respectively; where $\varepsilon_{True,t}$ and $\varepsilon_{fict.,t}$ denote differences of martingale sequence; and where $\mathcal{V}^{1/2}(S,I,\theta)$ is such that $\big(\mathcal{V}^{1/2}(S,I,\theta)\big)\big(\mathcal{V}^{1/2}(S,I,\theta)\big)' = \mathcal{V}(S,I,\theta)$,
with the function $\mathcal{V}$ defined in (\ref{eq:VarSvector}). To capture the idea that the two countries are affected by very similar standardized shocks, which are the $\varepsilon_{True,t}$'s and the $\varepsilon_{fict.,t}$'s, we further assume that:
$$
\mathbb{V}ar_t \left(
\left[\begin{array}{c}
\varepsilon_{True,t} \\
\varepsilon_{fict.,t} \\
\end{array}\right]
\right) \approx
\left[\begin{array}{cc}
{\bf Id} & {\bf Id}\\
{\bf Id} & {\bf Id}
\end{array}\right].
$$
On top of the transition equation (\ref{eq:transitCounter}), the state-space model comprehends the following measurement equation:
\begin{equation}\label{eq:measurCounter}
D_t^{obs.} = D_t,
\end{equation}
where $D_t^{obs.}$ is the observed vector of numbers of deaths (in the ``observed'' country). By construction, the filtered variables $X_t$ resulting from this augmented state-space framework are exactly equal the the ones of the regular state-space. Indeed, it produces the moment $\mathbb{E}(X_t|D_t^{obs},D_{t-1}^{obs},\dots)$, which is the same in both state-space models. However, $X_t^*$ will be different from $X_t$ since the implemented policies are different, and their impacts are non-trivial since they are non-linearly propagated in the state-space. 

As a last step, we provide backward path estimates using the Bryson-Frazier smoother and compare the paths of $D_t$ and $D_t^*$ produced by the smoother. We used Bryson-Frazier rather than Rauch-Tung-Striebel because the latter requires to invert the conditional variance-covariance matrices of the transition equations, which are of size $(9N\times 9N)$. In our empirical application (51 states), this results in matrices of size $(459\times 459)$ that have to be inverted for each day of data. This results in a large numerical instability. Instead, the Bryson-Frasier smoother only requires the inversion of the variance-covariance matrix of the observables, that is of matrices of size $(N\times N)$. 



\section{Infected populations in travel ban counterfactual} \label{sec:flows}

Figure \ref{fig:flows} showcases the posterior (smoothed)  mean of the cumulative number of infected individuals in the baseline scenario for some select states. The figure also shows the smoothed mean of the cumulative number of infected individuals in the counterfactual in which a federal interstate travel ban goes into effect on March 17, 2020.

\section{Sensitivity analysis} \label{app:sens}

We run several analyses to understand how sensitive our results are with respect to changes in our parameter values. Table \ref{tbl:fed} shows confidence bands that are derived from re-estimating the number of preventable deaths if the policy that deviated from the data was either twice or half as impactful. What we mean in precise terms by this is that if, for example, we assume that a counterfactual is carried out with respect to changes in a mask mandate, we would evaluate the number of preventable deaths by assuming that $\theta_{M}^{low}$ is either half or twice as large as indicated in Table \ref{tbl:est}. This sensitivity analysis provides confidence bands for our estimates of the number of deaths that could have been prevented by adopting policies different than the ones that were adopted in reality by considering that the policies may have a different impact on reducing transmission rates and traveler and commuter inflows than what we assume in our study.

We also carry out additional sensitivity analyses with respect to two key parameters in our model: the number of days that it takes for an infection to resolve, and the fatality rate of the disease. Tables \ref{tbl:fed_10days} through \ref{tbl:fed_highdelta} repeat the experiments of Table \ref{tbl:fed} by assuming that it either takes 10 or 20 days for an infection to resolve, or that the fatality rate of the disease is 0.4\% or 1\%. 
To obtain the results, we re-estimate the parameters $\beta$, $\sigma$, and $\rho$ that would be necessary under the new assumptions for the fatality rate or the number of days for infection resolution. The sensitivity analyses are carried out one-by-one.

We find that adopting early federal containment policies would have been less impactful and prevented less deaths if an infection took longer time to resolve. Vice versa, we find that early federal action would have been more impactful if an infection took less time to resolve. This is primarily because, if the disease were less severe than we assumed and it took less time to resolve an infection, then the methodology estimates that there must have been many more infections early on in the sample to match the number of deaths observed throughout the sample. That means that early action would have been more impactful if the disease is less severe than we assumed. Note that this is always a conclusion based on matching the amount of death cases that we observe in the data. If the disease were less severe and we still observed as many deaths as in the data, then the only possible explanation is that there must have been more infections early on in the sample. As a result, early action would have been the more impactful.

We find that the number of preventable deaths would not be much different if the virus were less lethal. However, we find that the costs of inaction at the state level would be much higher if the virus were more lethal. We find that close to 2,000,000 additional deaths would have been observed if no state enacted any policy and if the lethality rate of the virus were 1\%. These observations further corroborate our findings of the success of the policies enacted by the individual states to prevent COVID-19 deaths both at the state and federal levels.

\section{Computational details}

\subsection{Conditional covariances for the single-state model}\label{App:singleState}

We have:
\begin{eqnarray}
\mathbb{E}_{t-1}(r_t + d_t) &=& \mathbb{E}_{t-1}(d_t + \mathbb{E}_{t-1}(r_t|d_t)) - (\gamma + \delta)I_{t-1} \\
\mathbb{V}ar_{t-1}(r_t + d_t) &=&  \mathbb{V}ar_{t-1}(d_t + \mathbb{E}_{t-1}(r_t|d_t))+\mathbb{E}_{t-1}(\mathbb{V}ar_{t-1}(d_t + r_t|d_t)) \nonumber \\
&=&  \mathbb{V}ar_{t-1}\left(\frac{1 -\delta-\gamma}{1-\delta}d_{t}\right)+\mathbb{E}_{t-1}\left((I_{t-1}-d_t)\frac{\gamma(1-\delta-\gamma)}{(1-\delta)^2}\right) \nonumber \\
&=&  \left(\frac{1 -\delta-\gamma}{1-\delta}\right)^2I_{t-1}\delta(1-\delta)+\mathbb{E}_{t-1}\left((I_{t-1}-d_t)\frac{\gamma(1-\delta-\gamma)}{(1-\delta)^2}\right) \nonumber \\
&=&  \underbrace{\frac{(1 -\delta-\gamma)^2\delta+\gamma(1-\delta-\gamma)}{1-\delta}}_{=: \nu} I_{t-1}. \label{eq:condVarsingle}
\end{eqnarray}

Because $d_t + r_t$, on the one hand, and $i_t$, on the other hand, are independent conditional on the information available on date $t-1$, it follows that:
\begin{eqnarray}
\mathbb{E}_{t-1}(\Delta I_t) &=& \frac{S_{t-1}}{N}I_{t-1} \beta_{t-1} - (\gamma + \delta) I_{t-1}\\
\mathbb{V}ar_{t-1}(\Delta I_t) &=& \frac{S_{t-1}}{N}I_{t-1} \beta_{t-1}  + \nu I_{t-1}.
\end{eqnarray}

{\bf Remark}: Let us introduce $Z_t \equiv \Delta I_t - \mathbb{E}_{t-1}(\Delta I_t)$. By construction, $\mathbb{E}_{t-1}(Z_t) = 0$. It can be seen that $Z_t$ is the sum of $I_{t-1}$ i.i.d. random variables. Hence, if $I_{t-1}$ is large, we approximately have:
$$
Z_t \sim \mathcal{N}\left(0, \frac{S_{t-1}}{N}I_{t-1} \beta_{t-1}  + \nu I_{t-1}\right),
$$
conditional on the information available on $t-1$. This is also true for $i_t$, $r_t$ and $d_t + r_t$. Because the conditional variances of these variables are in $I_{t-1}$ (and not in $I_{t-1}^2$), it follows that, when $I_{t-1}$ is large, the deterministic version of the SIR model provides a good approximation of the dynamics of $(S,I,R,D)$ -- at least up to potential stochastic variation of $\beta_t$.\\

In the remaining of this appendix, we detail the computation of some of the covariances appearing in equation\,(\ref{eq:cov_single})
\begin{eqnarray*}
\mathbb{C}ov_{t-1}(\varepsilon_{D,t},\varepsilon_{R,t}) &=& \mathbb{C}ov_{t-1}(r_t,d_t)= \mathbb{C}ov_{t-1}(\mathbb{E}_{t-1}(r_t|d_t),d_t) \\
&=& \mathbb{C}ov_{t-1}\left(\frac{(I_{t-1}-d_t)\gamma}{1-\delta},d_t\right) \\
&=& - \frac{\gamma}{1-\delta}\mathbb{V}ar_{t-1}(d_t) = -\gamma\delta I_{t-1}.
\end{eqnarray*}

\begin{eqnarray*}
\mathbb{C}ov_{t-1}(\varepsilon_{D,t},\varepsilon_{I,t}) &=& -\mathbb{C}ov_{t-1}(d_t,d_t+r_t)= -\left(\mathbb{E}_{t-1}(d_t(d_t+r_t))-\mathbb{E}_{t-1}(d_t)\mathbb{E}_{t-1}(d_t+r_t)\right)\\
&=& - \mathbb{V}ar_{t-1}(d_t) - \mathbb{C}ov_{t-1}(r_t,d_t)\\
&=& - \delta(1-\delta)I_{t-1} + \gamma\delta I_{t-1}\\
&=& - \delta(1-\delta-\gamma)I_{t-1}.
\end{eqnarray*}

\begin{eqnarray*}
\mathbb{V}ar_{t-1}\left(r_t\right)&=& \mathbb{V}ar_{t-1}(\mathbb{E}_{t-1}(r_t|d_t)) + \mathbb{E}_{t-1}(\mathbb{V}ar_{t-1}(r_t|d_t)) \\
&=& \mathbb{V}ar_{t-1}\left(\frac{(I_{t-1}-d_t)\gamma}{1-\delta}\right) + \mathbb{E}_{t-1}\left((I_{t-1}-d_t)\frac{\gamma(1-\delta-\gamma)}{(1-\delta)^2}\right) \\
&=& \frac{\gamma^2\delta}{1-\delta}I_{t-1} + \frac{\gamma(1-\delta-\gamma)}{1-\delta}I_{t-1} = (\gamma - \gamma^2)I_{t-1}.
\end{eqnarray*}

\begin{eqnarray*}
\mathbb{C}ov_{t-1}(\varepsilon_{R,t},\varepsilon_{I,t}) &=& -\mathbb{C}ov_{t-1}(r_t,d_t+r_t)= -\left(\mathbb{E}_{t-1}(r_t(d_t+r_t))-\mathbb{E}_{t-1}(r_t)\mathbb{E}_{t-1}(d_t+r_t)\right)\\
&=& - \mathbb{V}ar_{t-1}(r_t) - \mathbb{C}ov_{t-1}(r_t,d_t)\\
&=& - (\gamma - \gamma^2)I_{t-1} + \gamma\delta I_{t-1}\\
&=& - \gamma(1 - \gamma-\delta)I_{t-1}.
\end{eqnarray*}

\subsection{Conditional means and variances in the multi-state model}\label{App:multiState}

\subsubsection{Conditional variance of commute and travel flows}

Let us denote by $\mbox{Flow}_{com,I,t}$ the vector whose $i^{th}$ component is the time-weighted net inflow of infected commuters in state $i$. (Remember that $\tau$ is the fraction of time spent by commuters in the state where they work.) We have:
	\begin{eqnarray*}
\mbox{Flow}_{com,I,t} &=& \tau\mbox{Flow}^{\leftarrow}_{com,I,t-1}-\tau\mbox{Flow}^{\rightarrow}_{com,I,t-1}.
	\end{eqnarray*}
	Using equations\,(\ref{eq:PhiSCleft}) to (\ref{eq:OmegaC}), we have, in particular:
	$$
	\mathbb{E}(\mbox{Flow}_{com,I,t}|I_t) = \tau \phi_{com,I,t}^{\leftarrow} - \tau\phi_{com,I,t}^{\rightarrow} = \Omega_{com,t}I_{t}.
	$$
	We  also have 
	$$
	\mathbb{C}ov(\mbox{Flow}_{com,I,k,t},\mbox{Flow}_{com,I,j,t}|I_t) =
	\left\{
	\begin{array}{ll}
	 - \tau^2w_{com,t}^{k,j}I_{t}^{k} - \tau^2 w_{com,t}^{j,k}I_{t}^{j} & \mbox{if $j \ne k$}\\
	 \tau^2 \left(\sum_{k \ne j}  w_{com,t}^{k,j} I_{t}^{k}\right) + \tau^2 \left(\sum_{j \ne k} w_{com,t}^{j,k}\right) I_{t}^{j} & \mbox{if $j = k$},
	 \end{array}
	 \right.
	$$
	that is, in vectorial form:
	\begin{eqnarray*}
	&&\mathbb{V}ar(\mbox{Flow}_{com,I,t}|I_t) =\\
	&& -\tau^2W_{com,t} \odot (I_t {\bf 1}') - \tau^2W_{com,t}' \odot ({\bf 1}{I_t}') + \mbox{\bf d}(\tau^2W_{com,t}'I_t+\tau^2(W_{com,t}{\bf 1})\odot I_t),
	\end{eqnarray*}
	which we denote by
	\begin{equation}\label{eq:VarFlowIC}
	\mathbb{V}ar(\mbox{Flow}_{com,I,t}|I_t) = \mathcal{C}(W_{com,t},I_t,\tau),
	\end{equation}
	where function $\mathcal{C}$ is defined by
	\begin{equation}\label{eq:functionC}
	 \mathcal{C}(W,Z,\tau) = \tau^2\left\{  - W \odot (Z {\bf 1}') -  W' \odot ({\bf 1}{Z}') + \mbox{\bf d}(W'Z+(W{\bf 1})\odot Z)\right\}.
	\end{equation}
	The same type of computation leads to
	\begin{eqnarray}\label{eq:VarFlowSC}
	\mathbb{V}ar(\mbox{Flow}_{com,S,t}|S_t) &=& \mathcal{C}(W_{com,t},S_t,\tau)\\
	\mathbb{V}ar(\mbox{Flow}_{trav,I,t}|I_t) &=& \mathcal{C}(W_{trav,t},I_t,1)\\
	\mathbb{V}ar(\mbox{Flow}_{trav,S,t}|S_t) &=& \mathcal{C}(W_{trav,t},S_t,1).
	\end{eqnarray}

\subsubsection{Conditional variance of new infections}

For any pair of independent random vectors $X$ and $Y$, we have
\begin{eqnarray*}
\mathbb{V}ar(X \odot Y) &=& \mathbb{E}(XX' )\odot \mathbb{E}(YY') - \mathbb{E}(X)\mathbb{E}(X)' \odot \mathbb{E}(Y)\mathbb{E}(Y)'\\
&=& \mathbb{V}ar(X)\odot \mathbb{V}ar(Y) + \mathbb{V}ar(X) \odot \mathbb{E}(Y)\mathbb{E}(Y)'+ \mathbb{V}ar(Y) \odot \mathbb{E}(X)\mathbb{E}(X)'.
\end{eqnarray*}

Therefore:
\begin{eqnarray*}
&&\mathbb{V}ar\Big((I_{t-1} + \mbox{Flow}_{com,I,t-1} + \mbox{Flow}_{trav,I,t-1}) \odot (S_{t-1} + \mbox{Flow}_{com,S,t-1}+ \mbox{Flow}_{trav,S,t-1})\Big|I_{t-1},S_{t-1}\Big)\\
&=& \mathbb{V}ar\Big(\mbox{Flow}_{com,I,t-1}+\mbox{Flow}_{trav,I,t-1}\Big|I_{t-1}\Big) \odot \mathbb{V}ar\Big(\mbox{Flow}_{com,S,t-1}+\mbox{Flow}_{trav,S,t-1}\Big|S_{t-1}\Big)+\\
&&   \mathbb{V}ar\Big(\mbox{Flow}_{com,I,t-1}+\mbox{Flow}_{trav,I,t-1}\Big|I_{t-1}\Big) \odot \Big(({\bf Id} + \Omega_{t-1})S_{t-1}S_{t-1}'({\bf Id} + \Omega_{t-1})'\Big) +\\
&&  \mathbb{V}ar\Big(\mbox{Flow}_{com,S,t-1}+\mbox{Flow}_{trav,S,t-1}\Big|S_{t-1}\Big) \odot \Big(({\bf Id} + \Omega_{t-1})I_{t-1}I_{t-1}'({\bf Id} + \Omega_{t-1})'\Big) \\
&=&  \Big(\mathcal{C}(W_{com,t-1},I_{t-1},\tau)+\mathcal{C}(W_{trav,t-1},I_{t-1},1)\Big) \odot  \Big(\mathcal{C}(W_{com,t-1},S_{t-1},\tau)+\mathcal{C}(W_{trav,t-1},S_{t-1},1)\Big) + \\
&& \Big(\mathcal{C}(W_{com,t-1},I_{t-1},\tau) + \mathcal{C}(W_{trav,t-1},I_{t-1},1)\Big)\odot \Big(({\bf Id} + \Omega_{t-1})S_{t-1}S_{t-1}'({\bf Id} + \Omega_{t-1})'\Big)  +\\
&& \Big(\mathcal{C}(W_{com,t-1},S_{t-1},\tau)+ \mathcal{C}(W_{trav,t-1},S_{t-1},1)\Big) \odot \Big(({\bf Id} + \Omega_{t-1})I_{t-1}I_{t-1}'({\bf Id} + \Omega_{t-1})'\Big) \\
&=:& \mathcal{D}(W_{com,t-1},W_{trav,t-1},S_{t-1},I_{t-1},\tau,\Omega_{t-1}),
\end{eqnarray*}
where $\Omega_{t-1} = \Omega_{trav,t-1}+ \Omega_{com,t-1}$ and where
\begin{eqnarray}\label{eq:mathcalD}
 \mathcal{D}(W_1,W_2,S,I,\tau,\Omega) &=& \Big(\mathcal{C}(W_1,I,\tau) + \mathcal{C}(W_2,I,1) \Big)\odot  \Big(\mathcal{C}(W_1,S,\tau)+ \mathcal{C}(W_2,S,1)\Big) + \nonumber\\
&& \Big(\mathcal{C}(W_1,I,\tau)+ \mathcal{C}(W_2,I,1) \Big) \odot \Big(({\bf Id} + \Omega)SS'({\bf Id} + \Omega)'\Big)  + \nonumber\\
&&\Big( \mathcal{C}(W_1,S,\tau)+ \mathcal{C}(W_2,S,1)\Big) \odot \Big(({\bf Id} + \Omega)II'({\bf Id} + \Omega)'\Big),
\end{eqnarray}
function $\mathcal{C}$ being defined in (\ref{eq:functionC}).\\

\subsubsection{Conditional variance of susceptibles}
Let us use the notation:
$$
I^*_t = I_{t} + \mbox{Flow}_{com,I,t} + \mbox{Flow}_{trav,I,t}\quad\mbox{and}\quad S^*_t = S_{t} + \mbox{Flow}_{com,S,t} + \mbox{Flow}_{trav,S,t}.
$$
Using the law of total variance:
\begin{eqnarray*}
&&\mathbb{V}ar_{t-1}\left(S_t\right)\\
&=& \mathbb{V}ar_{t-1}\left(\mathbb{E}_{t-1}(S_t|I_{t-1},S_{t-1},\mbox{Flow}_{com,I,t-1},\mbox{Flow}_{com,S,t-1},\mbox{Flow}_{trav,I,t-1},\mbox{Flow}_{trav,S,t-1})\right) +\\
&& \mathbb{E}_{t-1}\left(\mathbb{V}ar_{t-1}(S_t|I_{t-1},S_{t-1},\mbox{Flow}_{com,I,t-1},\mbox{Flow}_{com,S,t-1},\mbox{Flow}_{trav,I,t-1},\mbox{Flow}_{trav,S,t-1})\right)\\
&=& \mathbb{V}ar_{t-1}\left(\theta_{\beta,t-1}\odot \beta_{t-1} \odot \frac{{\bf 1}}{{\bf p}}\odot I^*_{t-1} \odot S^*_{t-1} \right) + \mathbb{E}_{t-1}\left(\mbox{\bf d}\left(\theta_{\beta,t-1}\odot \beta_{t-1} \odot \frac{{\bf 1}}{{\bf p}}\odot I^*_{t-1} \odot S^*_{t-1}\right)\right)\\
&=& \left[\left(\theta_{\beta,t-1}\odot \beta_{t-1} \odot \frac{{\bf 1}}{{\bf p}}\right)\left(\theta_{\beta,t-1}\odot \beta_{t-1} \odot \frac{{\bf 1}}{{\bf p}}\right)'\right] \odot \mathcal{D}(W_{com,t-1},W_{trav,t-1},S_{t-1},I_{t-1},\tau,\Omega_{t-1}) + \\
&& \mbox{\bf d}\left(\theta_{\beta,t-1}\odot \beta_{t-1} \odot \frac{{\bf 1}}{{\bf p}}\odot [({\bf Id} + \Omega_{t-1})I_{t-1}] \odot [({\bf Id} + \Omega_{t-1})S_{t-1}]\right),
\end{eqnarray*}
Let's denote the previous conditional variance by $ \Theta_{t-1}$. We have:
\begin{eqnarray} \label{eq:Theta}
\Theta_{t-1} &=& \mathbb{V}ar_{t-1}\left(S_t\right)  \nonumber\\
&:=& \left[\left(\theta_{\beta,t-1}\odot \beta_{t-1} \odot \frac{{\bf 1}}{{\bf p}}\right)\left(\theta_{\beta,t-1}\odot \beta_{t-1} \odot \frac{{\bf 1}}{{\bf p}}\right)'\right] \odot \mathcal{D}(W_{com,t-1},W_{trav,t-1},S_{t-1},I_{t-1},\tau,\Omega_{t-1}) + \nonumber\\
&& \mbox{\bf d}\left(\theta_{\beta,t-1}\odot \beta_{t-1} \odot \frac{{\bf 1}}{{\bf p}}\odot [({\bf Id} + \Omega_{t-1})I_{t-1}] \odot [({\bf Id} + \Omega_{t-1})S_{t-1}]\right),
\end{eqnarray}
where $\Omega_{t-1} = \Omega_{com,t-1}+\Omega_{trav,t-1}$ and function $\mathcal{D}$ is defined by (\ref{eq:mathcalD}).

\subsubsection{Conditional variance of the state vector}\label{app:subCondiVar}

What precedes implies that:
\begin{eqnarray}\label{eq:VarSvector}
&& \mathcal{V}(S_{t-1},I_{t-1}):= \mathbb{V}ar_{t-1}\left(
\left[
\begin{array}{c}
D_{t}\\
S_{t}\\
I_{t}\\
R_{t}\\
\beta_t
\end{array}
\right]
\right) = \\
&&
\left[
\begin{array}{ccccc}
\delta(1-\delta)\mbox{\bf d}(I_{t-1}) & {\bf 0} & -\delta(1-\delta-\gamma)\mbox{\bf d}(I_{t-1}) & -\delta\gamma\mbox{\bf d}(I_{t-1}) & {\bf 0}\\
{\bf 0} & \Theta_{t-1}& -\Theta_{t-1} & {\bf 0}  & {\bf 0}\\
-\delta(1-\delta-\gamma)\mbox{\bf d}(I_{t-1})& -\Theta_{t-1}& \Theta_{t-1}  + \nu \mbox{\bf d}(I_{t-1}) & - \gamma(1 - \gamma-\delta)\mbox{\bf d}(I_{t-1}) & {\bf 0}\\
 -\delta\gamma\mbox{\bf d}(I_{t-1}) & {\bf 0} & - \gamma(1 - \gamma-\delta)\mbox{\bf d}(I_{t-1}) & (\gamma - \gamma^2)\mbox{\bf d}(I_{t-1}) & {\bf 0} \\
 {\bf 0}  & {\bf 0}  & {\bf 0}  & {\bf 0}  & \Omega_{\beta,t-1} 
\end{array}
\right],\nonumber
\end{eqnarray}
with
\begin{eqnarray*}
\Omega_{\beta,t-1} &=&  \sigma^2\Delta t \mbox{\bf d}(\sqrt{\beta_{t-1}})\cdot \Sigma \cdot \mbox{\bf d}(\sqrt{\beta_{t-1}}),
\end{eqnarray*}
and where $\Theta_t$ is defined by equations\,(\ref{eq:Theta}), and $\nu$ is defined in equation\,(\ref{eq:condVarsingle}).


\subsection{Jacobian matrix for extended Kalman filter implementation}\label{app:Jacobian_filter}
	
We hereby provide the formulas for the Jacobian computation in the extended Kalman filter recursions. Denoting by $J = \partial \mathbb{E}_{t-1}[D_t,\, S_t,\, I_t,\, R_t,\, \beta_t^0]/\partial [D_{t-1},\, S_{t-1},\, I_{t-1},\, R_{t-1},\, \beta_{t-1}^0]$, we have:

\begin{eqnarray}
\notag &&J =  \left[
	\begin{array}{ccccc}
	{\bf Id} & 0 & \delta {\bf Id} & 0 & 0\\
	0 & {\bf Id} & 0 & 0 & 0 \\
	0 & 0 & (1-\delta -\gamma){\bf Id}  & 0 & 0 \\
	0 & 0 & \gamma {\bf Id} & {\bf Id} & 0 \\
	0 & 0 & 0 & 0 & (1 - \kappa){\bf Id}
	\end{array}
	\right] + \\
	&&  \quad 
\left[
	\begin{array}{c}
	0 \\
	-{\bf Id}\\
	+{\bf Id}\\
	0\\
	0
	\end{array}
	\right] \frac{\partial\left(\theta_{S,t-1}\odot \theta_{M,t-1}  \odot \beta_{t-1}^0 \odot \frac{\bf 1}{\bf p} \odot ([{\bf Id} + \Omega_{t-1}] I_{t-1} )\odot ([{\bf Id} + \Omega_{t-1}] S_{t-1} )  \right)}{\partial\, [D_{t-1},\, S_{t-1},\, I_{t-1},\, R_{t-1},\, \beta_{t-1}^0]}.
\end{eqnarray}
The matrix of partial derivatives is given by:
\begin{eqnarray}
	\notag && \frac{\partial\left(\theta_{S,t-1}\odot \theta_{M,t-1}  \odot \beta_{t-1}^0 \odot \frac{\bf 1}{\bf p} \odot ([{\bf Id} + \Omega_{t-1}] I_{t-1} )\odot ([{\bf Id} + \Omega_{t-1}] S_{t-1} )  \right)}{\partial\, [D_{t-1},\, S_{t-1},\, I_{t-1},\, R_{t-1},\, \beta_{t-1}^0]}\\
	 &=& \left[ \begin{array}{c}
	 	0\\
	 	\left(\mathrm{\mathbf{ Id}} + \Omega_{t-1}\right)\times \mathrm{\mathbf{d}}\left(\theta_{S,t-1}\odot \theta_{M,t-1}  \odot \beta_{t-1}^0 \odot \frac{\bf 1}{\bf p}\odot ([{\bf Id} + \Omega_{t-1}] I_{t-1} )\right)\\
	 	\left(\mathrm{\mathbf{ Id}} + \Omega_{t-1}\right)\times \mathrm{\mathbf{d}}\left(\theta_{S,t-1}\odot \theta_{M,t-1}  \odot \beta_{t-1}^0 \odot \frac{\bf 1}{\bf p}\odot ([{\bf Id} + \Omega_{t-1}] S_{t-1} )\right)\\
	 	0\\
	 	\mathrm{\mathbf{d}}\left(\theta_{S,t-1}\odot \theta_{M,t-1}  \odot \frac{\bf 1}{\bf p} \odot ([{\bf Id} + \Omega_{t-1}] I_{t-1} )\odot ([{\bf Id} + \Omega_{t-1}] S_{t-1} )   \right)
	 \end{array}\right]'.
\end{eqnarray}

\begin{sidewaystable}[p]
\centering
\begin{tabularx}{0.9\textwidth}{|X | c c c | c |}
\hline
\multirow{2}{*}{Counterfactual assumption} & \multicolumn{3}{c|}{Federal mandate active on:} & Deaths in excess \\
& Stay-at-home & Mask & Travel ban & of baseline \\
\hline\hline
Strict stay-at-home orders, mask mandates, and & 3/20 & 4/17 & 3/17 & -180,595 \\
travel bans for all states. & & & &  \footnotesize [-215,046; -137,288] \\
Strict stay-at-home orders in all states, and all other & 3/20 & & & -129,202 \\
state-level policies remain as in the data. & & & & \footnotesize [-181,352; -77,566]\\
Strict mask mandates in all states, and all other state-level & & 4/17 & &  -141,048 \\
policies remain as in the data. & & & & \footnotesize [-116,362; -167,026] \\
Strict travel bans in all states, and all other state-level & & & 3/17 & -8,149 \\
policies remain as in the data. & & & & \footnotesize [-8,773; -4,121]\\
\hline\hline
No stay-at-home order, mask mandate, or travel ban & & & & +1,044,270 \\
in any state. & & & &  \footnotesize [+458,475; +1,521,927]\\
No stay-at-home order in any state, but all other & & & & +615,569 \\
policies remain as in the data. & & & &  \footnotesize [+229,183; +1,429,593] \\
No mask mandate in any states, but all other & & & & +922,910 \\
policies remain as in the data. & & & & \footnotesize [+264,408; +1,342,444] \\
No travel ban in any states, but all other & & & & +3,346 \\
policies remain as in the data. & & & &  \footnotesize [+1,511; +3,596] \\
\hline\hline
Strict travel bans in all states by February 12, 2020, while  & & & 2/12 &  -9,487 \\
all other state-level policies remain as in the data. & & & & \footnotesize [-10,186; -4,788] \\
\hline
\end{tabularx}
\caption{Results of the counterfactual experiments in which we assume that all states jointly deviate from their enacted policies and adopt either strict or loose versions of the policies instead. The reported values are excess deaths relative to the number of U.S. deaths recorded in our data on November 30, 2020. In the counterfactuals, we compute the trajectories of death counts per state under the alternative policy scenarios that are consistent with the transmission rates filtered from the observed data.
The values in brackets give confidence bounds based on a sensitivity analysis of the estimates of the impact of the different policies on transmission rates and traveler and commuter inflow. The lower bounds assume that any policy that deviates from what it was in the data is half as impactful, while the upper bound assumes that any policy that deviates is twice as impactful. Table \ref{tbl:est} in Appendix \ref{app:sens} provides the parameter values used for the sensitivity analyses.
In the sensitivity analysis, we proceed in a similar way as for the counterfactuals and first compute posterior means of the state-level transmission rates that would explain the observed death counts under the assumption of alternative effectiveness for the different policies. We then compute the number of death that would have been observed if the policies had changed while keeping the recomputed trajectories of the transmission rates fixed. } 
\label{tbl:fed}
\end{sidewaystable}

\begin{table}[tp]
\centering
\begin{tabular}{|c r|| c  r | | c  r |}
\hline
$\gamma$ & $0.07$ & $\delta$ & $0.0004$ & $\kappa$ & $0.001$ \\
& $[0.05, 0.1]$ &  & $[0.0003, 0.0007]$ &  &  \\
\hline
$\beta$ & $0.16$ & $\sigma$ & $0.05$ & $\rho$ & $0.49$ \\
\hline
$\theta_S^{low}$ & $0.64$ & $\theta_T^{low}$ & $0.10$ & $\theta_M^{low}$ & $0.58$ \\
 & $[0.32, 0.82]$ & & $[0.05, 0.55]$ &  & $[0.29, 0.79]$ \\
\hline
$\tau_{com}$ & $0.36$ & & & $\tau_{trav}$ & $4.00$ \\
\hline
\end{tabular}
\caption{Parameter values. The values in bracket give alternative parametrizations used for a sensitivity analysis of our results in Appendix \ref{app:sens}.}
\label{tbl:est}
\end{table}

\begin{sidewaystable}[tp]
\centering
\begin{tabularx}{0.9\textwidth}{|X | c c c | c |}
\hline
\multirow{2}{*}{Counterfactual assumption ($\gamma = 0.1$)} & \multicolumn{3}{c|}{Federal mandate active on:} & Deaths in excess \\
& Stay-at-home & Mask & Travel ban & of baseline \\
\hline\hline
Strict stay-at-home orders, mask mandates, and & 3/20 & 4/17 & 3/17 & -190,481 \\
travel bans for all states. & & & &  \\
Strict stay-at-home orders in all states, and all other & 3/20 & & & -137,919 \\
state-level policies remain as in the data. & & & & \\
Strict mask mandates in all states, and all other state-level & & 4/17 & &  -156,990 \\
policies remain as in the data. & & & & \\
Strict travel bans in all states, and all other state-level & & & 3/17 & -8,456 \\
policies remain as in the data. & & & & \\
\hline\hline
No stay-at-home order, mask mandate, or travel ban & & & & +1,024,136 \\
in any state. & & & & \\
No stay-at-home order in any state, but all other & & & & +662,202 \\
policies remain as in the data. & & & & \\
No mask mandate in any states, but all other & & & & +991,286 \\
policies remain as in the data. & & & & \\
No travel ban in any states, but all other & & & & +3,647 \\
policies remain as in the data. & & & &  \\
\hline\hline
Mask mandate in all states on March 19, 2020, and all other &  & 3/20 & & -235,479 \\
state-level policies remain as in the data. & & & & \\
\hline\hline
Strict travel bans in all states by February 12, 2020, while  & & & 2/12 &  -9,610 \\
all other state-level policies remain as in the data. & & & & \\
\hline
\end{tabularx}
\caption{Results of the counterfactual experiments in which we assume that all states jointly deviate from their enacted policies and adopt either strict or loose versions of the policies instead. 
Here, we assume that it takes on average 10 days for an infection to resolve, while keeping the fatality rate of the disease fixed at 0.6\%.
The reported values are excess deaths relative to the number of U.S. deaths recorded in our data on November 30, 2020. In the counterfactuals, we compute the trajectories of death counts per state under the alternative policy scenarios that are consistent with the transmission rates filtered from the observed data.
The values in brackets give confidence bounds based on a sensitivity analysis of the estimates of the impact of the different policies on transmission rates and traveler and commuter inflow. The lower bounds assume that any policy that deviates from what it was in the data is half as impactful, while the upper bound assumes that any policy that deviates is twice as impactful. Table \ref{tbl:est} in Appendix \ref{app:sens} provides the parameter values used for the sensitivity analyses.
In the sensitivity analysis, we proceed in a similar way as for the counterfactuals and first compute posterior means of the state-level transmission rates that would explain the observed death counts under the assumption of alternative effectiveness for the different policies. We then compute the number of death that would have been observed if the policies had changed while keeping the recomputed trajectories of the transmission rates fixed. } 
\label{tbl:fed_10days}
\end{sidewaystable}

\begin{sidewaystable}[tp]
\centering
\begin{tabularx}{0.9\textwidth}{|X | c c c | c |}
\hline
\multirow{2}{*}{Counterfactual assumption ($\gamma = 0.05$)} & \multicolumn{3}{c|}{Federal mandate active on:} & Deaths in excess \\
& Stay-at-home & Mask & Travel ban & of baseline \\
\hline\hline
Strict stay-at-home orders, mask mandates, and & 3/20 & 4/17 & 3/17 & -165,664 \\
travel bans for all states. & & & &  \\
Strict stay-at-home orders in all states, and all other & 3/20 & & & -119,630 \\
state-level policies remain as in the data. & & & & \\
Strict mask mandates in all states, and all other state-level & & 4/17 & &  -114,864 \\
policies remain as in the data. & & & & \\
Strict travel bans in all states, and all other state-level & & & 3/17 & -8,136 \\
policies remain as in the data. & & & & \\
\hline\hline
No stay-at-home order, mask mandate, or travel ban & & & & +1,096,115 \\
in any state. & & & & \\
No stay-at-home order in any state, but all other & & & & +648,965 \\
policies remain as in the data. & & & & \\
No mask mandate in any states, but all other & & & & +717,271 \\
policies remain as in the data. & & & & \\
No travel ban in any states, but all other & & & & +2,943 \\
policies remain as in the data. & & & &  \\
\hline\hline
Strict travel bans in all states by February 12, 2020, while  & & & 2/12 &  -5,610 \\
all other state-level policies remain as in the data. & & & & \\
\hline
\end{tabularx}
\caption{Results of the counterfactual experiments in which we assume that all states jointly deviate from their enacted policies and adopt either strict or loose versions of the policies instead. 
Here, we assume that it takes on average 20 days for an infection to resolve, while keeping the fatality rate of the disease fixed at 0.6\%.
The reported values are excess deaths relative to the number of U.S. deaths recorded in our data on November 30, 2020. In the counterfactuals, we compute the trajectories of death counts per state under the alternative policy scenarios that are consistent with the transmission rates filtered from the observed data.
The values in brackets give confidence bounds based on a sensitivity analysis of the estimates of the impact of the different policies on transmission rates and traveler and commuter inflow. The lower bounds assume that any policy that deviates from what it was in the data is half as impactful, while the upper bound assumes that any policy that deviates is twice as impactful. Table \ref{tbl:est} in Appendix \ref{app:sens} provides the parameter values used for the sensitivity analyses.
In the sensitivity analysis, we proceed in a similar way as for the counterfactuals and first compute posterior means of the state-level transmission rates that would explain the observed death counts under the assumption of alternative effectiveness for the different policies. We then compute the number of death that would have been observed if the policies had changed while keeping the recomputed trajectories of the transmission rates fixed. } 
\label{tbl:fed_20days}
\end{sidewaystable}

\begin{sidewaystable}[tp]
\centering
\begin{tabularx}{0.9\textwidth}{|X | c c c | c |}
\hline
\multirow{2}{*}{Counterfactual assumption ($\delta = 0.0003$)} & \multicolumn{3}{c|}{Federal mandate active on:} & Deaths in excess \\
& Stay-at-home & Mask & Travel ban & of baseline \\
\hline\hline
Strict stay-at-home orders, mask mandates, and & 3/20 & 4/17 & 3/17 & -171,375 \\
travel bans for all states. & & & & \\
Strict stay-at-home orders in all states, and all other & 3/20 & & & -116,363 \\
state-level policies remain as in the data. & & & &\\
Strict mask mandates in all states, and all other state-level & & 4/17 & &  -136,004 \\
policies remain as in the data. & & & & \\
Strict travel bans in all states, and all other state-level & & & 3/17 & -8,450 \\
policies remain as in the data. & & & & \\
\hline\hline
No stay-at-home order, mask mandate, or travel ban & & & & +660,986 \\
in any state. & & & & \\
No stay-at-home order in any state, but all other & & & & +424,480 \\
policies remain as in the data. & & & & \\
No mask mandate in any states, but all other & & & & +594,679 \\
policies remain as in the data. & & & & \\
No travel ban in any states, but all other & & & & +2,791 \\
policies remain as in the data. & & & &  \\
\hline\hline
Strict travel bans in all states by February 12, 2020, while  & & & 2/12 &  -5,057 \\
all other state-level policies remain as in the data. & & & & \\
\hline
\end{tabularx}
\caption{Results of the counterfactual experiments in which we assume that all states jointly deviate from their enacted policies and adopt either strict or loose versions of the policies instead. 
Here, we assume that the fatality rate of the disease is 0.4\% instead of 0.6\%, and keep the number of days that it take for an infection to resolve at 14 days. 
The reported values are excess deaths relative to the number of U.S. deaths recorded in our data on November 30, 2020. In the counterfactuals, we compute the trajectories of death counts per state under the alternative policy scenarios that are consistent with the transmission rates filtered from the observed data.
The values in brackets give confidence bounds based on a sensitivity analysis of the estimates of the impact of the different policies on transmission rates and traveler and commuter inflow. The lower bounds assume that any policy that deviates from what it was in the data is half as impactful, while the upper bound assumes that any policy that deviates is twice as impactful. Table \ref{tbl:est} in Appendix \ref{app:sens} provides the parameter values used for the sensitivity analyses.
In the sensitivity analysis, we proceed in a similar way as for the counterfactuals and first compute posterior means of the state-level transmission rates that would explain the observed death counts under the assumption of alternative effectiveness for the different policies. We then compute the number of death that would have been observed if the policies had changed while keeping the recomputed trajectories of the transmission rates fixed. } 
\label{tbl:fed_lowdelta}
\end{sidewaystable}

\begin{sidewaystable}[tp]
\centering
\begin{tabularx}{0.9\textwidth}{|X | c c c | c |}
\hline
\multirow{2}{*}{Counterfactual assumption ($\delta = 0.0007$)} & \multicolumn{3}{c|}{Federal mandate active on:} & Deaths in excess \\
& Stay-at-home & Mask & Travel ban & of baseline \\
\hline\hline
Strict stay-at-home orders, mask mandates, and & 3/20 & 4/17 & 3/17 & -185,874 \\
travel bans for all states. & & & &  \\
Strict stay-at-home orders in all states, and all other & 3/20 & & & -138,178 \\
state-level policies remain as in the data. & & & & \\
Strict mask mandates in all states, and all other state-level & & 4/17 & &  -140,718 \\
policies remain as in the data. & & & & \\
Strict travel bans in all states, and all other state-level & & & 3/17 & -8,936 \\
policies remain as in the data. & & & & \\
\hline\hline
No stay-at-home order, mask mandate, or travel ban & & & & +1,803,895 \\
in any state. & & & & \\
No stay-at-home order in any state, but all other & & & & +910,766 \\
policies remain as in the data. & & & & \\
No mask mandate in any states, but all other & & & & +1,530,444 \\
policies remain as in the data. & & & & \\
No travel ban in any states, but all other & & & & +3,788 \\
policies remain as in the data. & & & & \\
\hline\hline
Strict travel bans in all states by February 12, 2020, while  & & & 2/12 &  -8,033 \\
all other state-level policies remain as in the data. & & & & \\
\hline
\end{tabularx}
\caption{Results of the counterfactual experiments in which we assume that all states jointly deviate from their enacted policies and adopt either strict or loose versions of the policies instead. 
Here, we assume that the fatality rate of the disease is 1\% instead of 0.6\%, and keep the number of days that it take for an infection to resolve at 14 days. 
The reported values are excess deaths relative to the number of U.S. deaths recorded in our data on November 30, 2020. In the counterfactuals, we compute the trajectories of death counts per state under the alternative policy scenarios that are consistent with the transmission rates filtered from the observed data.
The values in brackets give confidence bounds based on a sensitivity analysis of the estimates of the impact of the different policies on transmission rates and traveler and commuter inflow. The lower bounds assume that any policy that deviates from what it was in the data is half as impactful, while the upper bound assumes that any policy that deviates is twice as impactful. Table \ref{tbl:est} in Appendix \ref{app:sens} provides the parameter values used for the sensitivity analyses.
In the sensitivity analysis, we proceed in a similar way as for the counterfactuals and first compute posterior means of the state-level transmission rates that would explain the observed death counts under the assumption of alternative effectiveness for the different policies. We then compute the number of death that would have been observed if the policies had changed while keeping the recomputed trajectories of the transmission rates fixed. } 
\label{tbl:fed_highdelta}
\end{sidewaystable}

\begin{sidewaysfigure}[htbp]
\centering
\includegraphics[width = 0.95\textwidth]{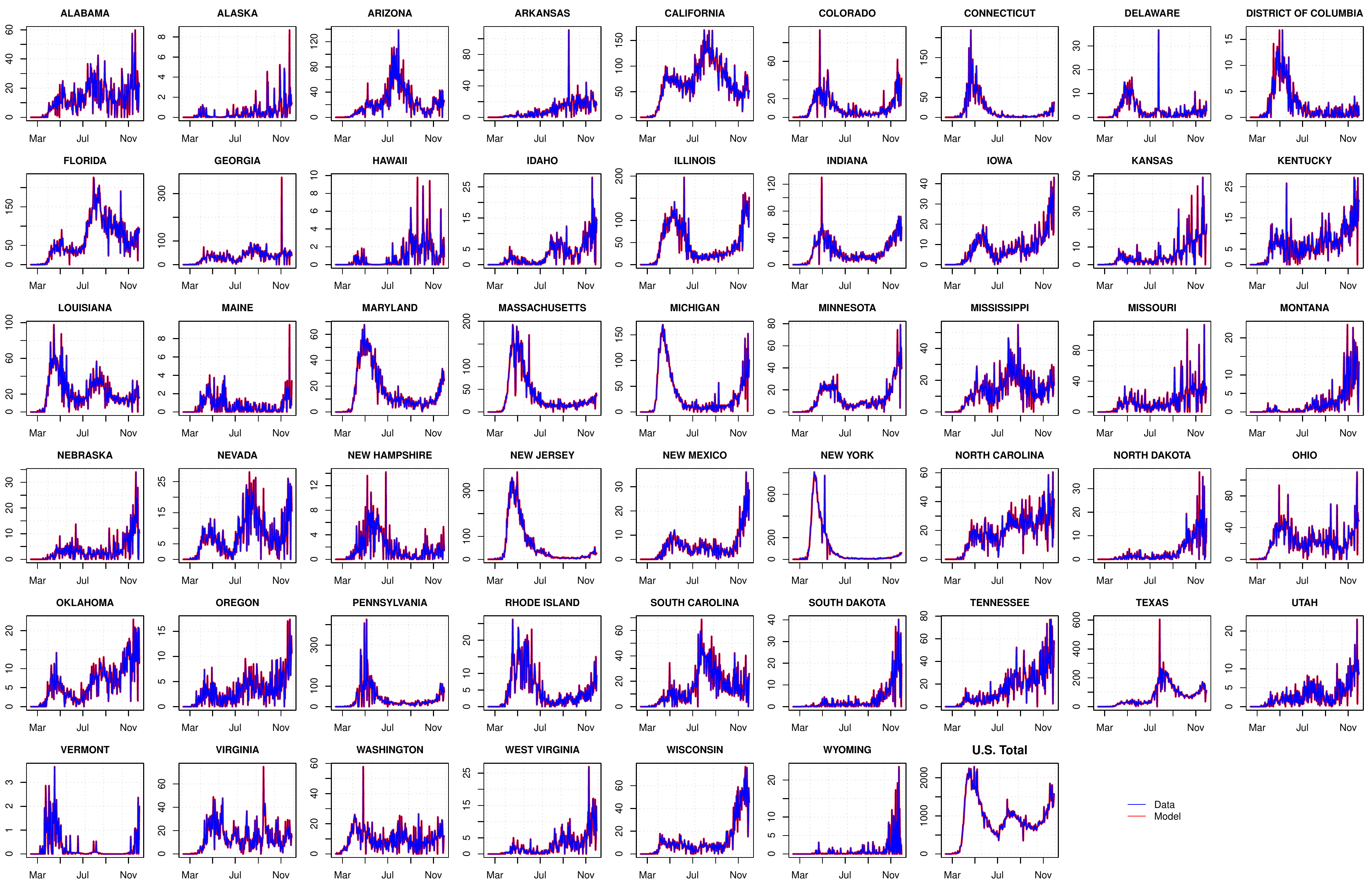}
\caption{Data and model-implied death counts across states. The model-implied death counts correspond to the number of fatalities that needed to have been recorded in our model to match the total number of deaths in a state at the end of the sample period (i.e., the smoother). Estimation is performed through forward/backward extended Kalman filtering, using the time series of death counts per state from February 12 to November 30, 2020. Our estimation methodology is detailed in Appendix \ref{sec:data}.}
\label{fig:death}
\end{sidewaysfigure}

\begin{sidewaysfigure}[htbp]
\centering
\includegraphics[width = 0.95\textwidth]{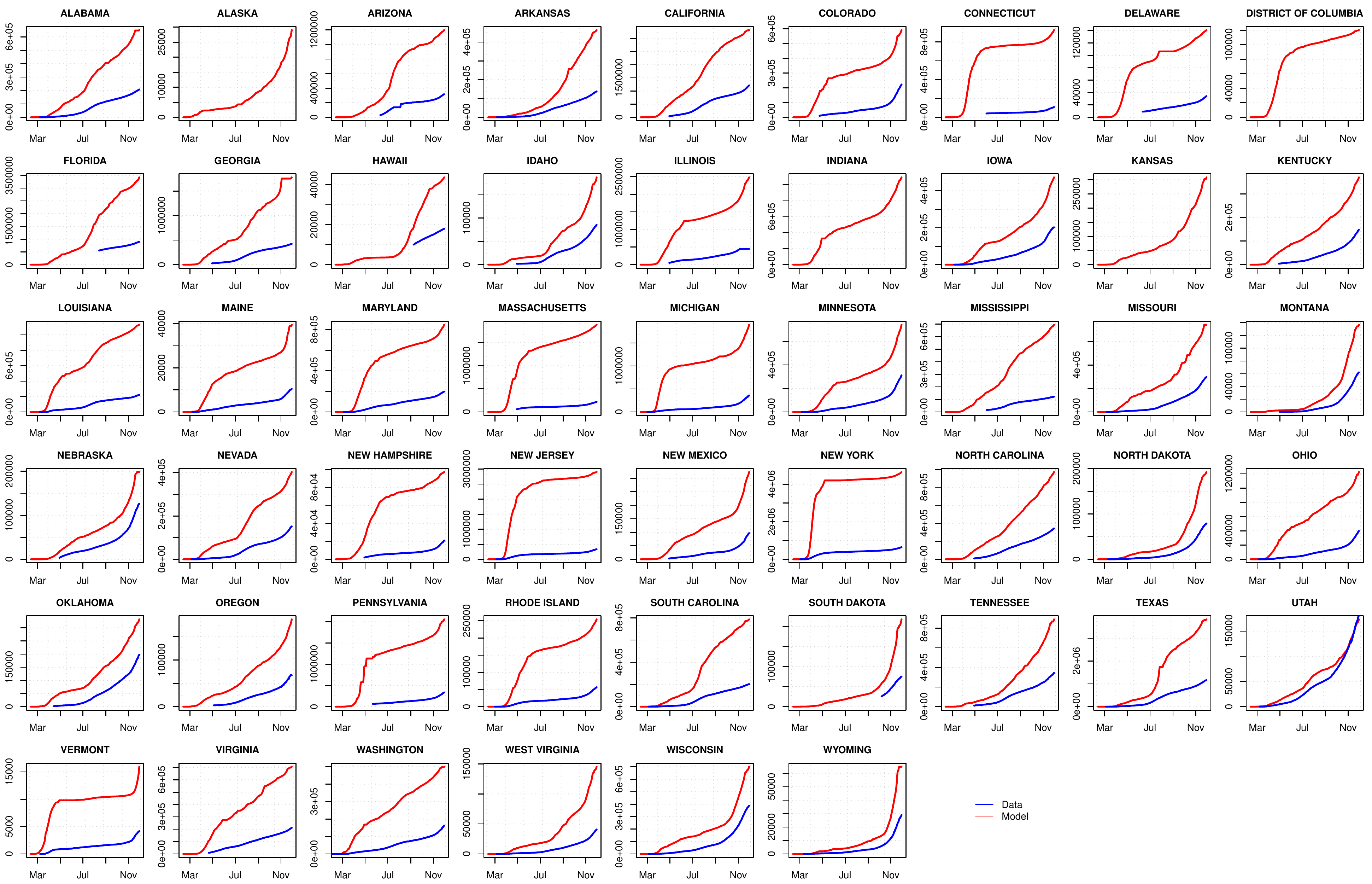}
\caption{Smoothed means of the model-implied cumulative number of COVID-19 infections in U.S. states. The figures also shows the number of infections that are recorded in state-level data from JHU. We obtain the cumulative number of infections by summing all populations per states except the susceptible. Estimation is performed through forward/backward extended Kalman filtering, using the time series of death counts per state from February 12 to November 30, 2020. Our estimation methodology is detailed in Appendix \ref{sec:data}.}
\label{fig:inf}
\end{sidewaysfigure}

\begin{figure}[t]
\begin{center}
\subfigure[South Carolina.]{
	\includegraphics[width=0.75\textwidth]{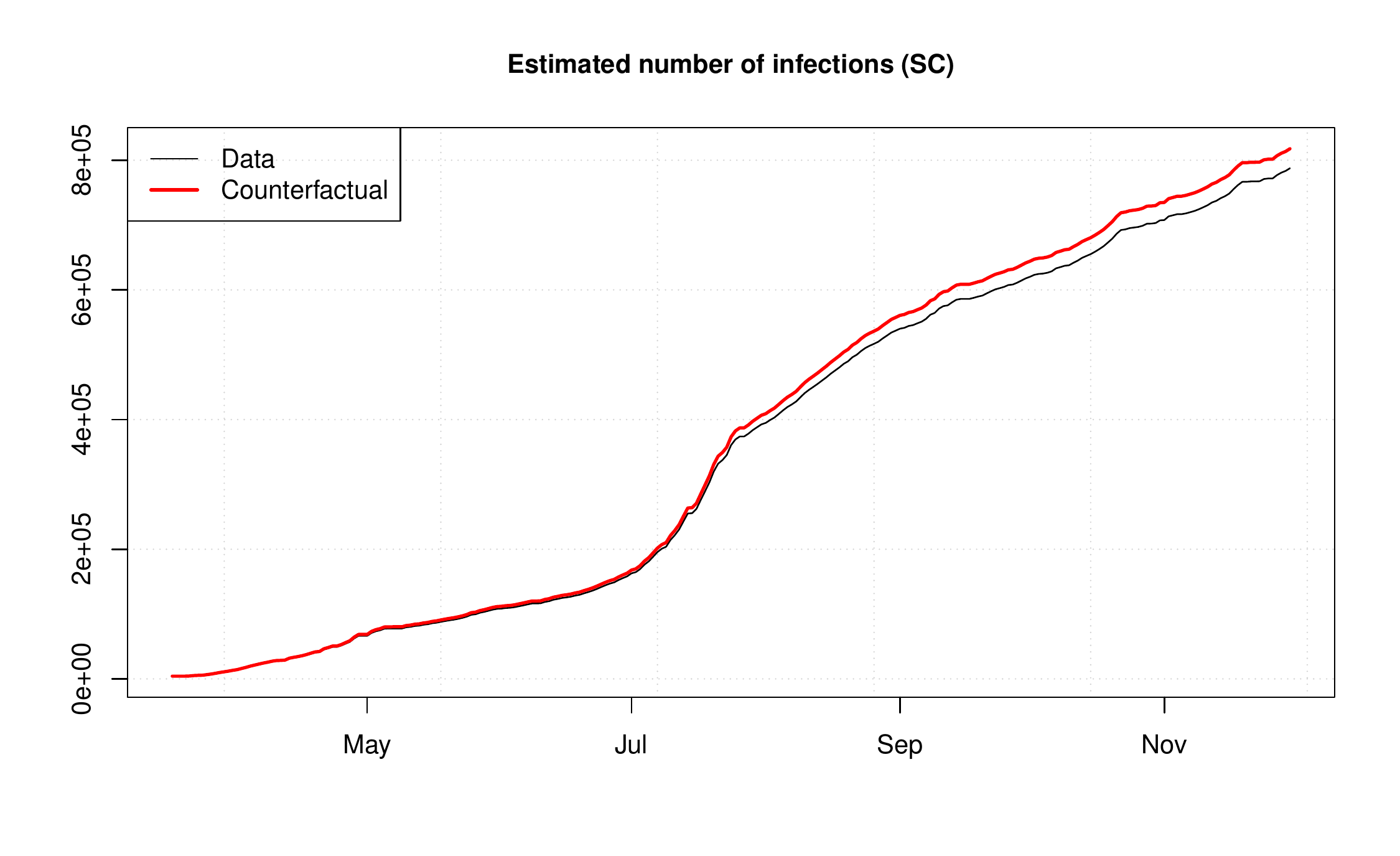}
} \\
\subfigure[Wyoming.]{
	\includegraphics[width=0.75\textwidth]{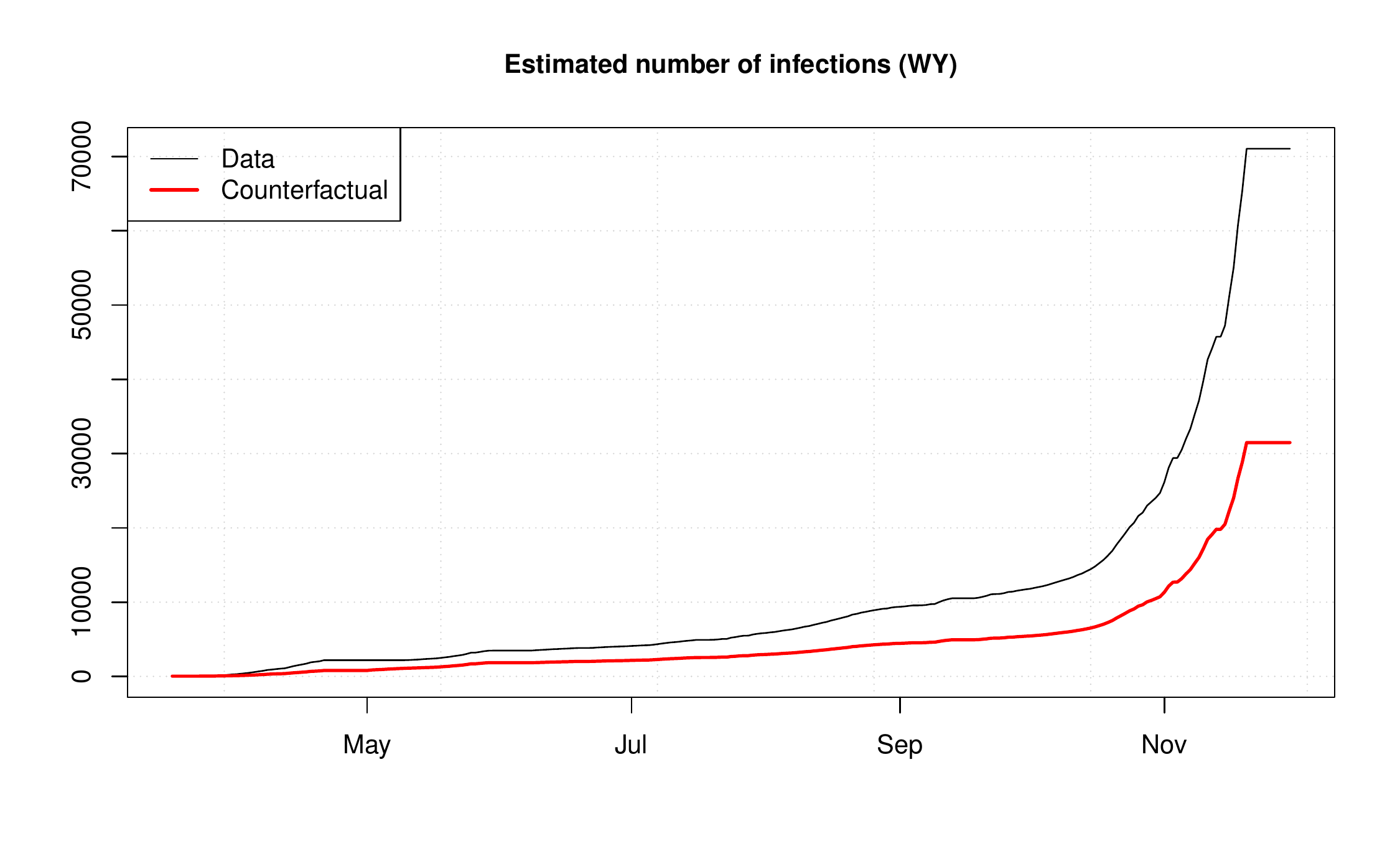}
} 
\end{center}
\caption{Time series of estimated number of infected individuals in select states. The grey line marks the posterior (smoothed) mean of the cumulative number of infections in the baseline. The red line denotes the posterior (smoothed) mean of the cumulative number of infections in the counterfactual in which a federal interstate travel ban goes into effect on March 17, 2020. Estimation is performed through forward/backward extended Kalman filtering, using the time series of death counts per state from February 12 to November 30, 2020. Our estimation methodology is detailed in Appendix \ref{sec:data}.}
\label{fig:flows}
\end{figure}

\end{document}